\documentclass[12pt]{article}
\usepackage{amsmath,epsfig,latexsym}

\def\bm#1{\mbox{\boldmath$#1$\unboldmath}}

\begin{document}

\begin{titlepage}

\begin{flushright}
CLNS~05/1949\\
December 17, 2005
\end{flushright}

\vspace{0.7cm}
\begin{center}
\bf
{\Large
Effective Field Theory and Heavy Quark Physics}\\[0.2cm]
(2004 TASI Lectures)
\end{center}

\vspace{0.8cm}
\begin{center}
{\sc Matthias Neubert}\\
\vspace{0.4cm}
{\sl $^a$\,Institute for High-Energy Phenomenology\\
Newman Laboratory for Elementary-Particle Physics, Cornell University\\
Ithaca, NY 14853, U.S.A.\\[0.3cm]
$^b$\,Institut f\"ur Theoretische Physik, Universit\"at Heidelberg\\
Philosophenweg 16, D--69120 Heidelberg, Germany}
\end{center}

\vspace{1.0cm}
\begin{abstract}
\vspace{0.2cm}
\noindent 
These notes are based on five lectures presented at the 2004 Theoretical 
Advanced Study Institute (TASI) on {\em ``Physics in $D\ge 4$''}. After a 
brief motivation of flavor physics, they provide a pedagogical introduction 
to effective field theory, the effective weak Lagrangian, and the technology
of renormalization-group improved perturbation theory. These general methods 
are then applied in the context of heavy-quarks physics, introducing the 
concepts of heavy-quark and soft-collinear effective theory.
\end{abstract}
\vfil

\end{titlepage}

\section{The Physics of Beauty}

Many of the unsolved problems in particle physics have their origin in the 
fact that we do not yet fully understand the properties of matter. In flavor 
physics, we study aspects of matter connected with the observation that its
fundamental constituents (quarks and leptons) come in replications known as 
generations. There exist some big, open questions in flavor physics, to which 
we would love to find some answers. Let me mention three of them:

\paragraph{What is the dynamics of flavor?}
The gauge forces in the Standard Model do not distinguish between fermions 
belonging to different generations. All charged leptons have the same 
electrical charge. All quarks carry the
same color charge. In almost all respects the fermions belonging to
different generations are equal -- but not quite, since
their masses are different. Today, we understand very little about the
underlying dynamics responsible for the phenomenon of generations.
Why do generations exist? Why are there three of them? Why are
the hierarchies of the fermion masses and mixing angles the way they are?
Why are these hierarchies different for quarks and leptons? We have
good reasons to expect that the answers to these questions,
if they can be found in the foreseeable future, will open the doors to 
some great discoveries, such as new symmetries, new forces, new dimensions, 
or something we have not yet thought about.

\paragraph{What is the origin of baryogenesis?}
The existential question about the origin of the matter-antimatter
asymmetry provides a link between particle physics and the evolution
of the Universe. The Standard Model satisfies the prerequisites for
baryogenesis as spelled out in the Sakharov criteria: baryon-number
violating processes are unsuppressed at high temperature;
CP-violating interactions are present due to complex couplings in
the quark (and presumably, the lepton) sector; non-equilibrium
processes can occur during phase transitions driven by the expansion
of the
Universe. However, quantitatively the observed matter abundance 
cannot be explained in the Standard Model (by many orders of
magnitude). Additional contributions, either due to new CP-violating
phases or new mechanisms of CP violation, are required. 

\paragraph{Are there connections between flavor physics and
TeV-scale physics?}
What can flavor physics tell us about the origin of electroweak
symmetry breaking? And, if the world is supersymmetric at some high
energy scale, what can flavor physics teach us about the mechanism of
Supersymmetry breaking? Virtually any extension of
the Standard Model that can solve the gauge hierarchy problem (i.e.,
the fact that the electroweak scale is so much lower than the GUT
scale) naturally contains a plethora of new flavor parameters. Some
prominent examples are:
\begin{itemize}
\item
Supersymmetry: hundreds of flavor- and/or CP-violating couplings, even in the
MSSM and its next-to-minimal variants
\item
extra dimensions: flavor parameters of Kaluza-Klein states
\item
Technicolor: flavor couplings of Techni-fermions
\item
multi-Higgs models: CP-violating Higgs couplings
\item
Little Higgs models: flavor couplings of new gauge bosons ($W'$, $Z'$)
and fermions ($t'$)
\end{itemize}
If New Physics exists at or below the TeV 
scale, its effects should show up, at some level of precision, in
flavor physics. Flavor- and/or CP-violating interactions can only be
studied using precision measurements at highest luminosity. In the future, 
such studies will profit from the fact that the relevant mass scales will
(hopefully) be known from the LHC.

To drive this last point home, let me recall some lessons from the 
past. Top quarks have been discovered through direct production at the
Tevatron. In that way, their mass, spin, and color charge have been 
determined. Accurate predictions for the mass were available before,
based on electroweak precision measurements at the $Z$ pole, but also
based on studies of $B$ mesons. The rates for $B$--$\bar B$ mixing, as
well as for rare flavor-changing neutral current (FCNC) processes such 
as $B\to X_s\gamma$, are very sensitive to the value of the top-quark 
mass. More importantly, everything else we know
about the top quark, such as its generation-changing couplings
$|V_{ts}|\approx 0.040$ and $|V_{td}|\approx 0.008$, as well as its
CP-violating interactions ($\mbox{arg}(V_{td})\approx -24^\circ$ with 
the standard choice of phase conventions), 
has come from studies of kaon and $B$ physics. 
Next, recall the example of neutrino oscillations. The existence of
neutrinos has been known for a long time, but it was the discovery of 
their flavor-changing interactions (neutrino oscillations) that has
revolutionized our thinking about the lepton sector. We have learned
that the hierarchy of the leptonic mixing matrix is very different
from that of the quark mixing matrix, and we have discovered that
leptogenesis and CP violation in the lepton sector may provide an
alternative mechanism for baryogenesis.

Exploring flavor aspects of the New Physics, whatever it
may be, is therefore not an exercise meant to fill the Particle Data
Book. Rather, it is of crucial relevance to answer some profound,
deep questions about Nature. Some questions for which we have a 
realistic chance of finding an answer with the help of a second-generation 
$B$-factory are:
\begin{itemize}
\item
Do non-standard CP phases exist? If so, this may provide new clues
about baryogenesis.
\item
Is the electroweak symmetry-breaking sector flavor blind (minimal
flavor violation)?
\item
Is the Supersymmetry-breaking sector flavor blind?
\item
Do right-handed currents exist? This may provide clues about new gauge
interactions and symmetries (left-right symmetry) at very high energy.
\end{itemize}
The interpretation of New Physics signals at present or future  
$B$-factories can be tricky. But since it is our hope to answer
some important questions, we must try as hard as we can. Flavor physics will
thus remain a valuable component in the 
comprehensive exploration of the TeV scale. It is a shame that this vital part
of physics is currently being terminated in the U.S.

\subsection*{Precision measurements in the quark sector}

At first sight, the presence of generations, i.e.\ the replication 
(triplication) of the fundamental fermions, makes the Standard Model more
ugly than it needed to be. The fermion masses and mixings constitute many of
the parameters of the Standard Model Lagrangian. Unlike the parameters in 
the gauge sector, these masses and mixings exhibit strongly hierarchical
patterns. Importantly, fermions of different generations can communicate via 
flavor-changing weak interactions. Indeed, these are the
{\em only\/} flavor-changing (i.e., generation-nondiagonal) interactions 
in the Standard Model. 

One of the main goals of the present $B$ factories -- 
one that has been achieved in a spectacular way! -- 
is the precise determination of the parameters of the
quark mixing matrix (the Cabibbo-Kobayashi-Maskawa matrix). 
Interestingly, the presence of at least three fermion generations allows for
CP violation to occur in flavor-changing weak decays, which is one of the 
prerequisites for an explanation of the matter-antimatter asymmetry observed
in the Universe. (Yet, as mentioned above, while baryogenesis is thus 
possible in the Standard Model, it has not been possible to explain the 
observed baryon asymmetry quantitatively.) 
Modern CKM physics is often described with the help of the unitarity-triangle 
relation 
\begin{equation}
   V_{ud} V_{ub}^* + V_{cd} V_{cb}^* + V_{td} V_{tb}^* = 0 \,,
\end{equation}
which contains the two smallest entries in the matrix, $V_{ub}$ and 
$V_{td}$.

Measuring the parameters of the unitarity triangle serves two purposes: 
first, to determine some fundamental parameters of the Standard Model, and 
secondly, to test whether the CKM mechanism 
of flavor and CP violation is indeed correct, or whether there are hints of
deviations from the Standard Model. By now, there is a plethora of ways in 
which the sides and angles of the unitarity triangle have been constrained
(see \cite{Charles:2004jd} for a comprehensive review). 
Until now no significant deviations from the CKM mechanism have been 
established. In particular, measurements of the sides of the triangle are
consistent with measurements of the angles. Also, the CP-violating phase 
of $V_{td}\sim e^{-i\beta}$, which is probed in particle-antiparticle 
mixing, is consistent with the phase of $V_{ub}\sim e^{-i\gamma}$, which 
is probed in rare exclusive $B$-meson decays.

Developing theoretical methods for a systematic analysis of exclusive 
hadronic decays such as $B\to\pi\pi$
has been one of the greatest challenges to heavy-flavor theory.
In the remaining lectures, I will discuss some of the theoretical concepts
relevant in this context.
Exclusive hadronic decays also feature prominently in searches for New
Physics at the $B$ factories, and indeed there exist some hints 
(not more) for possible deviations from the Standard Model in the measurements 
of CP asymmetries in some loop-dominated processes such as $B\to\phi K_S$ 
and $B\to\eta' K_S$. If confirmed, this could 
point toward the existence of some new FCNC 
transitions, which can effectively compete with Standard Model loop
amplitudes. This would be the first signal we have for new TeV-scale physics,
and it should provide enough motivation for you to follow the rest of these 
lectures.

\section{Effective Field Theory}

Effective field theory (EFT) is a very powerful tool in quantum field theory 
\cite{Polchinski:1992ed}. It 
provides a systematic formalism for the analysis of multi-scale problems. 
This is particularly important in QCD, where the value of the running coupling
$\alpha_s(\mu)$ can change significantly between different energy scales. As 
such, EFT greatly simplifies practical calculations in field theory; indeed, 
it often makes such calculations feasible. As we will discuss, 
EFT also provides a new, modern meaning to ``renormalization''.

The main idea of EFT is simply stated: Consider a quantum field theory with a
large, fundamental scale $M$. This could be the mass of a heavy particle, or 
some large (Euclidean) momentum transfer. Suppose we 
are interested in physics at energies $E$ (or momenta $p$) 
much smaller than $M$. 
How can we expand scattering or decay amplitudes in powers of $E/M$? The answer
to this question proceeds in several steps:
\begin{enumerate}
\item
Choose a cutoff $\Lambda<M$ and divide the fields of the theory into
low-frequency and high-frequency modes,
\begin{equation}
   \phi = \phi_L + \phi_H \,,
\end{equation}
where $\phi_L$ contains the Fourier modes with frequency $\omega<\Lambda$, 
while $\phi_H$ contains the remaining 
modes with frequency $\omega>\Lambda$. We can 
think of the cutoff as a ``threshold of ignorance'' in the sense that we 
may pretend to know nothing about the theory for scales above $\Lambda$ 
(which is indeed often the case). By construction, low-energy physics is 
described in terms of the $\phi_L$ fields. Everything we ever wish to know
about the theory (Feynman diagrams, scattering amplitudes, cross sections, 
decay rates, etc.) can be derived from vacuum correlation functions of these 
fields. These correlators can be obtained using 
\begin{equation}\label{correl}
   \langle 0|\,T\{ \phi_L(x_1)\dots\phi_L(x_n) \}\,0\rangle
   = \frac{1}{Z[0]} \left( -i\,\frac{\delta}{\delta J_L(x_1)} \right) 
   \dots \left( -i\,\frac{\delta}{\delta J_L(x_n)} \right)
   Z[J_L] \Big|_{J_L=0} \,,
\end{equation}
where
\begin{equation}
   Z[J_L] = \int{\cal D}\phi_L\,{\cal D}\phi_H\,
   e^{iS(\phi_L,\phi_H) + i\int d^Dx\,J_L(x)\,\phi_L(x)}
\end{equation}
is the generating functional of the theory. Here
$S(\phi_L,\phi_H)=\int d^Dx\,{\cal L}(x)$ is the action, $D$ is the dimension
of space-time, and we have only included sources $J_L$ for the light fields, 
as this suffices to compute the correlation functions in (\ref{correl}).
\item
In the next step, we perform the path integral over the high-frequency 
fields. This yields
\begin{equation}
   Z[J_L]\equiv \int{\cal D}\phi_L\,
   e^{iS_\Lambda(\phi_L) + i\int d^Dx\,J_L(x)\,\phi_L(x)} \,,
\end{equation}
where
\begin{equation}
   e^{iS_\Lambda(\phi_L)} = \int{\cal D}\phi_H\,e^{iS(\phi_L,\phi_H)}
\end{equation}
is called the ``Wilsonian effective action''. Note that, by construction, this 
action depends on the choice of the cutoff $\Lambda$ used to define the split
between low- and high-frequency modes. $S_\Lambda$ is non-local on scales 
$\Delta x^\mu\sim 1/\Lambda$, because high-frequency fluctuations have been 
removed from the theory. The process of removing these modes is often referred 
to as ``integrating out'' the high-frequency fields in the functional 
integral.
\item
In the final step, we expand the non-local action functional in terms of 
local operators composed of light fields. This process is 
called the (Wilsonian) operator-product expansion (OPE).
This expansion is possible because 
$E\ll\Lambda$ by assumption. The result can be cast in the form
\begin{equation}
   S_\Lambda(\phi_L) = \int d^Dx\,{\cal L}_\Lambda^{\rm eff}(x) \,,
\end{equation}
where
\begin{equation}\label{OPE}
   {\cal L}_\Lambda^{\rm eff}(x) = \sum_i\,g_i\,Q_i(\phi_L(x)) \,.
\end{equation}
This object is called the ``effective Lagrangian''. It is an infinite sum over
local operators $Q_i$ multiplied by coupling constants $g_i$, which are 
referred to as Wilson coefficients. In general, all operators allowed by the
symmetries of the theory are generated in the construction of the effective 
Lagrangian and appear in this sum. 
\end{enumerate}

Since there is always an infinite number of such  
operators, the question arises: How can the effective low-energy theory be 
predictive?
This is where the simple, but powerful trick of ``naive dimensional analysis'' 
comes to play. As is common practice 
in high-energy physics, let us work in 
units where $\hbar=c=1$. Then $[m]=[E]=[p]=[x^{-1}]=[t^{-1}]$ are all 
measured in the same units. We denote by $[g_i]=-\gamma_i$ the mass dimension 
of the effective couplings $g_i$. It follows that
\begin{equation}
   g_i = C_i\,M^{-\gamma_i}
\end{equation}
with dimensionless coefficients $C_i$. Since by assumption there is only a 
single fundamental scale $M$ in the theory, we expect that $C_i=O(1)$. This 
assertion is known as the hypothesis of ``naturalness''. Unless there is a
specific mechanism that could explain the smallness of the dimensionless 
numbers $C_i$, we should assume those numbers to be of $O(1)$. The presence of 
unusually large (e.g.\ $10^6$) or small (e.g.\ $10^{-6}$) numbers in a theory 
would appear ``unnatural'' and call for further explanation. 

At low energy ($E\ll\Lambda<M$), 
the contribution of a given operator $Q_i$ in the effective Lagrangian to an
observable (which for simplicity we assume to be dimensionless) is 
expected to scale as
\begin{equation}
   C_i \left( \frac{E}{M} \right)^{\gamma_i}
   = \begin{cases}
    O(1) \,; & \mbox{if $\gamma_i=0$,} \\
    \ll 1 \,; & \mbox{if $\gamma_i>0$,} \\
    \gg 1 \,; & \mbox{if $\gamma_i<0$.} 
   \end{cases}
\end{equation}
It follows that only operators whose couplings have $\gamma_i\le 0$ are 
important at low energy. This very fact is what makes the OPE 
a useful tool. Depending on the precision goal, one may 
truncate the series in (\ref{OPE}) at a given order in $E/M$. Once this is
done, only a finite (often small) number of operators $Q_i$ and couplings 
$g_i$ need to be retained. 

Let us go through the above arguments once again, being slightly more careful. 
Assuming weak coupling,\footnote{Interactions can change the naive scaling 
dimensions $\gamma_i$, as we will see later. For this reason, the $\gamma_i$ 
are referred to as ``anomalous dimensions''.}
we can use the free action to assign a scaling 
behavior with $E$ to all fields and couplings in the low-energy effective 
theory. Consider scalar $\phi^4$ theory as an example. The action is
\begin{equation}\label{phi4}
   S = \int d^D x \left( \frac12\,\partial_\mu\phi\,\partial^\mu\phi
   - \frac{m^2}{2}\,\phi^2 - \frac{\lambda}{4!}\,\phi^4 \right) .
\end{equation}
Using that $x\sim E^{-1}$ and $\partial^\mu\sim E$, and requiring that the 
action scale like $O(1)$ (in units of $\hbar$), we see that 
$\phi\sim E^{\frac{D}{2}-1}$. If we denote by 
$\delta_i$ the mass dimension of an operator $Q_i$, then 
$\gamma_i=\delta_i-D$. For the operators in the Lagrangian (\ref{phi4}) we 
find:
\begin{center}
\begin{tabular}{c|c|c|c}
 & $\delta_i$ & $\gamma_i$ & Coupling \\
\hline
$\partial_\mu\phi\,\partial^\mu\phi$ & $D$ & 0 & 1 \\
$\phi^4$ & $2D-4$ & $D-4$ & $\lambda\sim\Lambda^{4-D}$ \\
$\phi^2$ & $D-2$ & $-2$ & $m^2\sim\Lambda^2$ \\
\end{tabular}
\end{center}
More generally, an operator with $n_1$ scalar fields and $n_2$ derivatives 
has
\begin{equation}
   \delta_i = n_1 \left( \frac{D}{2} - 1 \right) + n_2 \,, \qquad
   \gamma_i = (n_1-2) \left( \frac{D}{2} - 1 \right) + (n_2-2) \,.
\end{equation}
It follows that for $D>2$ only few operators have $\gamma_i\le 0$.

A summary of these considerations is presented in Table~\ref{tab:termi}.
The common terminology of ``relevant'', ``marginal'', and ``irrelevant'' 
operators given there is without a doubt one of the worst misnomers is the 
history of physics. Really, ``relevant'' operators are usually unimportant, 
because they are forbidden by a symmetry (else they are disastrous, see 
below). ``Marginal'' operators are all there is in renormalizable quantum 
field theories. And ``irrelevant'' operators are those that are really 
interesting, because they teach us something about physics at the fundamental
scale $M$.

\begin{table}
\centerline{\parbox{14cm}{\caption{\label{tab:termi}
Classification of operators and couplings in the effective Lagrangian}}}
\vspace{0.1cm}
\begin{center}
\begin{tabular}{|c|c|c|}
\hline
Dimension & Importance for $E\to 0$ & Terminology \\
\hline
$\delta_i<D$, $\gamma_i<0$ & grows & relevant operators \\
 & & (super-renormalizable) \\
$\delta_i=D$, $\gamma_i=0$ & constant & marginal operators \\
 & & (renormalizable) \\
$\delta_i>D$, $\gamma_i>0$ & falls & irrelevant operators \\
 & & (non-renormalizable) \\
\hline
\end{tabular}
\end{center}
\end{table}

A crucial insight, which one may term the ``theorem of modesty'', is that no
quantum field theory is ever complete at arbitrarily high energy. 
At best it is an
EFT valid up to some cutoff scale $\Lambda$. This ``scale of ignorance'' is 
often a physical scale, such as the mass of a new particle, which has not yet
been discovered. When interpreted that way, many theories we know and love 
can be seen as EFTs: 
\begin{center}
\begin{tabular}{c|c|c}
High-energy theory & Fundamental scale & Low-energy theory \\
\hline
Standard Model & $M_W\sim 80$\,GeV & Fermi theory \\
GUT & $M_{\rm GUT}\sim 10^{16}$\,GeV & Standard Model \\
String theory & $M_S\sim 10^{18}$\,GeV & QFT \\
11-dim.\ $M$ theory & \dots & String theory \\
\dots & \dots & \dots \\
\end{tabular}
\end{center}

The arguments just presented provide a new perspective on 
renormalization. Instead of a paradigm of renormalizable theories based 
on the concept of systematic ``cancellations of infinities'', we should 
adopt the following, more physical point of view: 
\begin{itemize}
\item
Low-energy physics depends on the short-distance structure of the fundamental 
theory via relevant and marginal couplings, and possibly through some 
irrelevant couplings provided measurements are sufficiently precise. 
\item
``Non-renormalizable'' interactions are not forbidden; on the contrary, 
irrelevant operators always contribute at some level of precision. Their 
effects are simply numerically suppressed if the fundamental scale $M$ is 
much larger than the typical energies achievable experimentally.
\item
These non-renormalizable, ``irrelevant'' interactions tell us something about 
the physics at the cutoff scale $\Lambda\sim M$. 
\end{itemize} 
A corrolary to the second item is that, 
at low energies, all EFTs are ``automatically'' renormalizable
quantum field theories, provided that the cutoff scale $\Lambda$ is large.

The comment about ``irrelevant'' interactions in the third item is very 
powerful, so let us illustrate it with two prominent examples: 
{\em i)} 
Early measurements of the magnitude and energy dependence of 
weak-interaction processes at low energy have indicated the relevance of a 
high mass scale $M\sim 100$\,GeV. This was instrumental in finding the correct
theory of the weak interactions. 
{\em ii)} 
The local gauge symmetries of the Standard Model allow us to write down a 
dimension-5 operator of the type $g\,\nu^T HH\nu$ with $g\sim 1/\Lambda$. 
After electroweak symmetry breaking, this operator 
gives rise to a neutrino Majorana
mass term $m_\nu\sim v^2/\Lambda$, where $v\sim 246$\,GeV is the
vacuum expectation value of the Higgs field. Seen as an EFT, the 
Standard Model thus predicts the existence of neutrino masses, even though 
there are no right-handed neutrino fields in the theory. 
The seasaw mechanism
provides an explicit example of how such a mass term might be realized in a 
more fundamental theory. But unless we forbid the dimension-5 operator by 
imposing a symmetry such a lepton-number conservation, the existence 
of neutrino masses is a generic prediction of the Standard Model. The fact 
that the observed neutrino masses imply $\Lambda\sim 10^{14}$\,GeV not far 
from the energy scale where the three gauge couplings approximately unify 
is a strong argument in favor of the idea of Grand Unification.

On the other hand, super-renormalizable terms in an effective Lagrangian are 
problematic. Consider as an example the operator $\phi^2$ in scalar $\phi^4$ 
theory (i.e., the mass term for the scalar field). In $D=4$ dimensions we
have $\delta_i=2$, $\gamma_i=-2$, and so we expect that $m^2\sim\Lambda^2$ by
virtue of the hypothesis of naturalness. Since such large fluctuations are 
indeed generated in the functional integral, we expect the mass of the scalar 
field to be enormous (assuming the cutoff scale is large). But this 
is a contradiction: the $\phi$ particle would be heavy, and so it would not be 
part of the low-energy effective theory. No experiment at low energy could
produce such a heavy particle. This reasoning leads to a new paradigm of 
``naturalness'': EFTs should be natural in the sense that all mass terms
are forbidden by symmetries. These symmetries must be broken at low
energy, since otherwise everything would be massless. In the Standard Model, 
the fact that the fundamental forces are derived from gauge interactions 
guarantees that the spin-1 bosons (photon, gluons, $W$ and $Z$ bosons) are 
massless, while the fact that the weak interactions act on chiral 
(left-handed) fermions forbids fermion mass terms. Some of these symmetries
are broken at the electroweak scale, and so mass terms are generated at that 
scale.\footnote{This reasoning explains why $M_W\sim M_Z\sim 100$\,GeV. 
However, the expectation that also the fermion masses are of order the weak 
scale only works for the top quark. This is known as the flavor puzzle.}
Scalar particles, in particular the Higgs boson of the Standard Model, are 
not protected by any symmetry. In the context of EFTs they are not
allowed in the low-energy effective Lagrangian. This leads to the notion that 
the Standard Model is not a consistent (better, natural) EFT. The Standard 
Model Higgs sector is thus not expected to be the correct theory of 
electroweak symmetry breaking.

There are several possible ways out of this dilemma. Two conventional routes
are to invoke a new symmetry (Supersymmetry) to protect scalar particles from
acquiring masses of order the cutoff scale, or to abandon the 
idea of a fundamental scalar and instead realize the Higgs boson as a bound 
state of some new, strong interaction (Technicolor). 
Alternatively, the Higgs boson can be
made light if it is identified with the (pseudo-) Goldstone modes of a 
spontaneously broken global symmetry (little Higgs models). More 
recently, enlightened by the fact that there is no indication whatsoever for 
physics beyond the Standard Model, some theorists have given up on the idea 
of naturalness, and unnatural (e.g., fine-tuned) theories such as Split 
Supersymmetry \cite{Arkani-Hamed:2004fb,Arkani-Hamed:2004yi} have received a 
lot of attention. In these models the 
explanation and fine-tuning of parameters such as a light scalar mass (or
the cosmological constant) is derived from anthropic reasoning 
\cite{Weinberg:1987dv,Agrawal:1997gf}.

Let me finish this lecture with some comments on renormalization and 
running couplings. First, note that quantum corrections 
can alter the naive scaling relations for the dimensions of operators and 
couplings, giving rise to ``anomalous dimensions'' (a better term would be 
``quantum dimensions''). In a weakly coupled theory, these anomalous 
dimensions can be calculated using perturbation theory. Lecture~\ref{lec:4} 
shows what they are good for. Next, while so far we 
thought of $\phi_H$ as describing some heavy-particle fields (such as the 
top quark, or the weak gauge bosons $W$, $Z$ in the Standard Model), 
it is important to realize that these 
fields also describe the high-frequency quantum fluctuations of light or 
massless fields. Now consider an EFT with only light fields, in which we lower
the cutoff scale $\Lambda$ by some small amount $\delta\Lambda$. This 
corresponds to integrating out just a few high-frequency modes in a small 
slice between $\Lambda$ and $\Lambda-\delta\Lambda$ in energy. Since the 
operators $Q_i(\phi_L)$ remain the same in such a situation (since no heavy
particles are integrated out), the effects of lowering the cutoff must be 
absorbed entirely by a change of the effective couplings $C_i(\Lambda)$ in 
the effective Lagrangian ${\cal L}_\Lambda^{\rm eff}$. This provides an 
intuitive understanding of why the effective couplings are ``running 
couplings'', whose values depend on the cutoff.

\section{Effective Weak Interactions}

\begin{figure}
\begin{center}
\includegraphics[width=0.55\textwidth]{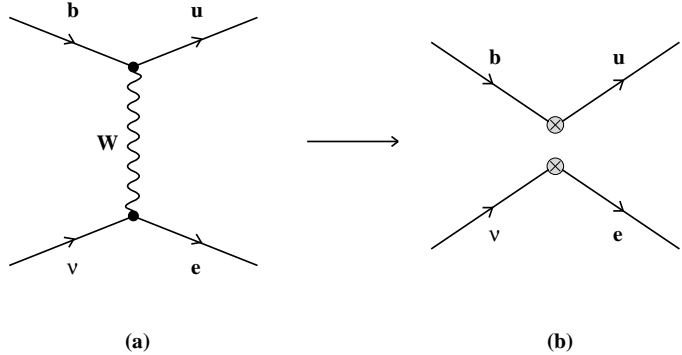} 
\end{center}
\vspace*{-0.5cm}
\caption{\label{fig:4fermi}
Example of an effective four-fermion interaction obtained by integrating out 
the $W$ boson in the Standard Model. The two crossed circles in the second 
graph represent a local four-quark operator in the effective theory. 
(Courtesy of A.~J.~Buras \cite{Buras:1998ra})}
\end{figure}

The couplings of the charged weak gauge bosons $W^\pm $ to fermions (quarks
and leptons) are the only flavor-changing interactions in the Standard Model.
When studying flavor-changing processes at low energy (i.e., $E\ll M_W$), we 
can integrate out the heavy bosons from the Standard Model Lagrangian. As 
illustrated in Figure~\ref{fig:4fermi}, this
gives rise to local four-fermion interactions. The resulting Fermi theory of 
weak interactions is particularly simple if we ignore the effects of QCD. 
Indeed, at tree level the path integral is Gaussian, and integrating over the 
$W^\pm $ fields gives the effective Lagrangian
\begin{equation}\label{Lweak}
   {\cal L}_{\rm weak}^{\rm eff} = - \frac{g^2}{8M_W^2} \left[
   J_\mu^-\,J^{+\mu} + \frac{1}{M_W^2}\,J_\mu^-\,(\partial^\mu\partial^\nu
   - g^{\mu\nu}\,\Box)\,J_\nu^+ + \dots \right] ,
\end{equation}
where $g^2/8M_W^2\equiv G_F/\sqrt2$ defines the Fermi constant 
($G_F=1.16639\cdot 10^{-5}$\,GeV$^{-2}$), and
\begin{equation}
   J_\mu^+ = V_{ij}\,\bar u_i\gamma_\mu(1-\gamma_5) d_j
    + \bar\nu_i\gamma_\mu(1-\gamma_5) l_i \,, \qquad
   J_\mu^- = (J_\mu^+)^\dagger
\end{equation}
are the charged currents. $V_{ij}$ are the elements of the 
CKM matrix, and a summation over flavor indices $i,j$ is understood.

Already the leading term in the effective weak Lagrangian (\ref{Lweak}) 
contains 
irrelevant interactions ($\delta_i=6$, $\gamma_i=2$), and indeed the coupling
constant $G_F\sim 1/M_W^2$ shows the expected suppression by two powers of 
the fundamental scale. Even before the discovery of the weak gauge bosons, 
experiments of low-energy weak interactions indicated that the fundamental 
scale of the weak force should be $(\sqrt2/G_F)^{1/2}\approx 110$\,GeV. It was
a triumph of particle physics when the heavy gauge bosons were subsequently
discovered at just that mass scale. The fact that there are no marginal 
operators in the effective weak Lagrangian explains the apparent 
``weakness'' of the weak interactions at low energy. On the contrary, at high
energy the weak force is unified with electromagnetism, and the two 
interactions are then governed by a single coupling constant. Subleading 
terms in the effective weak Lagrangian have dimension $\delta_i=8$ and higher.
Their effects are further suppressed by powers of $(E/M_W)^2$ and are tiny.
They can be neglected for (almost) all practical purposes.

Beyond the tree approximation, the question arises of how to account for the 
effects of the strong interactions in the derivation of the effective 
Lagrangian \cite{Buras:1998ra}. 
They are due to the high-frequency modes of the quark and
gluon fields. (The low-energy modes of these fields remain part of the 
low-energy theory, which still contains QCD.) Two problems arise: first, the
path integral is no longer Gaussian once QCD effects are taken into account; 
secondly, the strong interactions are not perturbative at low energy due to 
the confinement of colored particles into hadrons. One deals with these
difficulties using a general procedure called ``matching'', which consists of
the following steps:
\begin{enumerate}
\item
List all possible gauge-invariant operators of a given dimension allowed by 
the symmetries and quantum numbers associated with a given problem. The 
dimension of the operators is determined by the accuracy goal of the 
calculation, but generally $\delta_i=6$ for calculations in flavor physics. 
\item
Write down the OPE for the effective 
Lagrangian with {\em undetermined\/} couplings $C_i$, in our case
\begin{equation}
   {\cal L}_{\rm weak}^{\rm eff}
   = - \frac{G_F}{\sqrt2} \sum_i\,C_i\,Q_i \,.
\end{equation} 
Note that the Wilson coefficients $C_i$ are process independent, i.e., the 
same coefficients arise in the calculation of many different weak-interaction 
amplitudes.
\item
Determine the values of the coefficients $C_i(\mu)$ such that\footnote{I have 
switched notation here, such that $\mu$ takes the role of the cutoff scale 
$\Lambda$. In practice, one almost always uses dimensional regularization to 
perform calculations in quantum field theory, whereas a hard Wilsonian 
cutoff $\Lambda$ is used in conceptual arguments.}
\begin{equation}
   {\cal A}_n = \langle f_n|\,{\cal L}_{\rm SM}\,|i_n\rangle
   \stackrel{!}{=} \sum_i\,C_i(\mu)\,\langle f_n|\,Q_i\,|i_n\rangle
   + \mbox{higher power corrections}
\end{equation}
to a given order in perturbation theory. We need to study as many such matrix 
elements as necessary to determine the coefficients $C_i$ once and forever. 
Note that the choice of matrix elements is not unique. 
\end{enumerate}

In a weakly coupled theory 
all calculations can be done using perturbation theory, but what if the 
theory becomes strongly coupled at low energy, such as QCD? The crucial point
is that we can still determine the $C_i$ perturbatively if the theory 
is weakly coupled at high energy (asymptotic freedom at short distances). 
This statement becomes obvious if we recall the construction of the Wilsonian
effective action. The Wilson coefficient functions arise from integrating out
high-frequency quantum fluctuations above the cutoff scale $\Lambda$. Below
the scale $\Lambda$ the effective theory is, by construction, equivalent to
the fundamental theory order by order in power counting. Differences arise
only at high energy, where the effective theory misses the high-frequency 
modes of the full theory. The corresponding contributions are absorbed into
the Wilson coefficients. As along as the theory is perturbative (weakly 
coupled) at and above the scale $\Lambda$ (or $\mu$), the Wilson coefficients
are calculable in perturbation theory.
It follows that the Wilson coefficients in the effective Lagrangian
are insensitive to any infrared (IR) physics -- unlike the amplitudes 
themselves. As a result, matching calculations can be done using arbitrary 
IR regulators and working with free quark and gluon states. Obviously, this 
is a great advantage for actual calculations in QCD.

\subsection*{Four-fermion processes}

To construct the leading terms in the effective weak Lagrangian we take note
of the following facts:
\begin{itemize}
\item
Four fermion fields already make dimension $\delta_i=6$, so no derivatives or 
extra fields are allowed at this order.
\item
Weak interactions only involve left-handed fermion fields.
\item
Chirality is preserved in strong-interaction processes, since 
we can set $m_q=0$ at leading power.
\item
For the quark bilinears $\bar\psi_L\Gamma\psi_L$ with $\Gamma$ an
element of the Dirac basis, only the possibility $\Gamma=\gamma_\mu$ remains.
\item
Operators must be gauge invariant (in particular, color singlets) and Lorentz
invariant.
\end{itemize}

\paragraph{Example 1:}
Consider semileptonic decays such as $\bar B^0\to\pi^+ e^-\bar\nu_e$,
which are based on the quark transition $b\to u\,e^-\bar\nu_e$. From such 
processes one can extract the CKM element $|V_{ub}|$. The unique dimension-6 
operator in the corresponding effective Lagrangian is
\begin{equation}
   \bar e_L\gamma_\mu\nu_L\,\bar u_L^i\gamma^\mu b_L^i
   \,\,\hat{=}\,\, \bar u_L^i\gamma_\mu\nu_L\,\bar e_L\gamma^\mu b_L^i \,,
\end{equation}
where the second form is related to the first one by a Fierz transformation. 
This operator is generated by integrating out the $W$ boson in 
Figure~\ref{fig:4fermi}.
Below, we will often omit color indices $i$ whenever they are contracted 
between neighboring quark fields. Tree-level matching yields
\begin{equation}
   {\cal L}_{\rm eff} = - \frac{4G_F}{\sqrt2}\,V_{ub}\,C_1(\mu)\,
   \bar e_L\gamma_\mu\nu_L\,\bar u_L\gamma^\mu b_L
\end{equation}
with $C_1=1+O(\alpha_s)$.

\paragraph{Example 2:}
A more interesting case is offered by hadronic decays such as 
$\bar B^0\to\pi^+ D_s^-$, which are based on the quark transition 
$b\to u\,\bar c\,s$. Similar arguments now allow two operators differing in 
their color structure. Specifically,
\begin{equation}\label{example2}
   {\cal L}_{\rm eff} = - \frac{4G_F}{\sqrt2}\,V_{cs}^*\,V_{ub}
   \left[ C_1(\mu)\,\bar s_L^j\gamma_\mu c_L^j\,\bar u_L^i\gamma^\mu b_L^i
   + C_2(\mu)\,\bar s_L^i\gamma_\mu c_L^j\,\bar u_L^j\gamma^\mu b_L^i \right]
   ,
\end{equation}
where $C_1=1+O(\alpha_s)$ and $C_2=O(\alpha_s)$ follow again from tree-level 
matching. Using a Fierz rearrangement, the second operator above can also be 
written as $\bar u_L^j\gamma_\mu c_L^j\,\bar s_L^i\gamma^\mu b_L^i$. Note
also that
\begin{equation}
   \bar s_L\gamma_\mu t_a c_L\,\bar u_L\gamma^\mu t_a b_L
   = \frac12\,\bar s_L^i\gamma_\mu c_L^j\,\bar u_L^j\gamma^\mu b_L^i
   - \frac{1}{2N_c}\,\bar s_L\gamma_\mu c_L\,\bar u_L\gamma^\mu b_L
\end{equation}
gives nothing new. Here $t_a$ are the generators of color SU(3).

\subsection*{Matching at one-loop order}

\begin{figure}
\begin{center}
\includegraphics[width=0.65\textwidth]{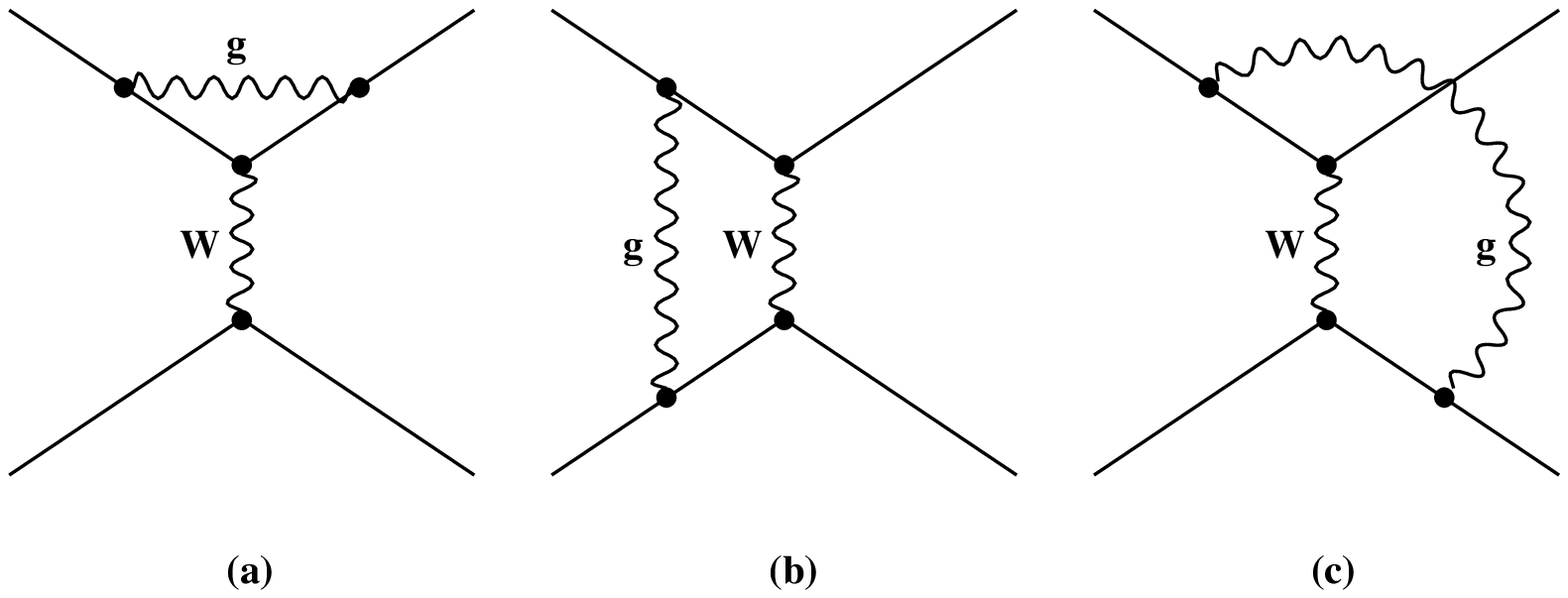} 
\includegraphics[width=0.65\textwidth]{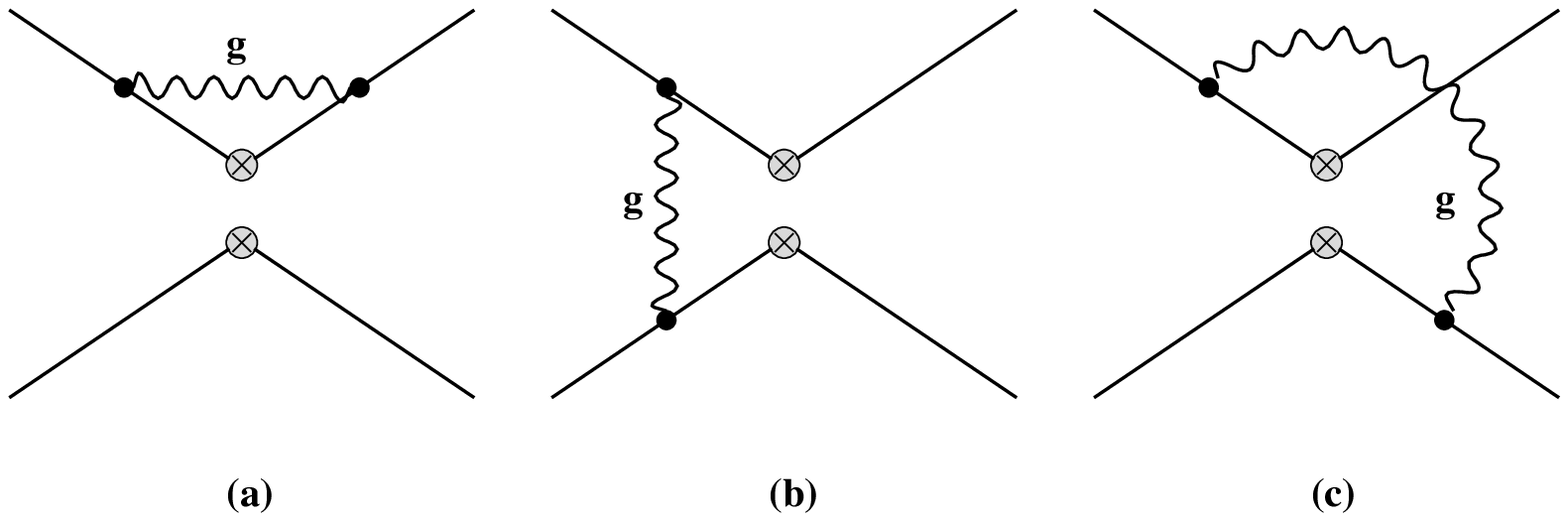} 
\end{center}
\vspace*{-0.5cm}
\caption{\label{fig:2}
One-loop QCD corrections to four-fermion weak-interaction amplitudes, both 
in the Standard Model (top row) and in the low-energy effective theory 
(bottom row). For $b\to u\,e^-\bar\nu_e$ only the diagrams labeled~(a) are
possible, whereas for $b\to u\,\bar c\,s$ all six diagrams contribute.
(Courtesy of A.~J.~Buras \cite{Buras:1998ra})}
\end{figure}

The one-loop QCD corrections to the $b\to u\,e^-\bar\nu_e$ transition in 
both the full theory and the effective theory are shown by the first 
diagram, labeled~(a), in each row in in Figure~\ref{fig:2}. (Diagrams~(b) 
and (c) are absent if the lower fermion line represents a lepton pair, as 
in the present case.) 
One finds that the radiative corrections in the two theories are identical, 
and so $C_1(\mu)=1$ to all orders in (QCD) perturbation theory. Note that in 
this example the Taylor expansion of the $W$-boson propagator trivially 
commutes with loop integrations.

The one-loop QCD corrections to the $b\to u\,\bar c\,s$ transition in both 
the full theory and the effective theory are represented by all six 
diagrams shown in Figure~\ref{fig:2}. 
The diagrams labeled~(b) and (c), which now contribute, are more interesting, 
and their evaluation
is considerably more difficult. One finds that in this example the two 
operations -- expansion of the $W$ propagator and integration over loop 
momenta -- do not commute. The reason is that in the last two diagrams in 
the top row the loop momentum flows through the $W$-boson propagator. Rather
than going through the full calculation in detail, we just note that
\begin{equation}
   \int d^Dp\,\frac{1}{M_W^2-p^2}\,f(p)
   \ne \frac{1}{M_W^2} \int d^Dp\,\left( 1 + \frac{p^2}{M_W^2} + \dots \right)
   f(p) \,.
\end{equation}
While the left-hand side is non-analytic in $M_W$, the right-hand side is 
obviously analytic. Differences between the two integrals arise from the 
region of large loop momenta where $p^2\sim M_W^2$. But for such large momenta
QCD is weakly coupled. Perturbation theory can thus be trusted to compute the 
differences between the matrix elements in the two theories, which are 
accounted for by the Wilson coefficient functions. Explicit
calculation of the diagrams in Figure~\ref{fig:2} gives (in the 
$\overline{\rm MS}$ subtraction scheme) \cite{Buchalla:1995vs}
\begin{eqnarray}\label{C1C2}
   C_1(\mu) &=& 1 + \frac{3}{N_c}\,\frac{\alpha_s(\mu)}{4\pi}
    \left( \ln\frac{M_W^2}{\mu^2} - \frac{11}{6} \right) + O(\alpha_s^2) \,,
    \nonumber\\
   C_2(\mu) &=& -3\,\frac{\alpha_s(\mu)}{4\pi}
    \left( \ln\frac{M_W^2}{\mu^2} - \frac{11}{6} \right) + O(\alpha_s^2) \,.
\end{eqnarray}
Some important comments are in order:
\begin{itemize}
\item
IR regulators (such as quark and gluon masses, external momenta, etc.) present
in intermediate steps of the calculation in both theories cancel in the 
results for the Wilson coefficients $C_i$.
\item
Matrix elements in the effective theory are often more singular than those in 
the full theory (which are ultraviolet (UV) finite in the present case) and 
require additional UV subtractions (operator renormalization). This gives rise 
to the renormalization-scale and -scheme dependence of the Wilson 
coefficients.
\item
The physical reason for this is that the mass $M_W$ acts as an UV regulator in
the box diagrams of the full theory. Roughly speaking, the logarithms in the
results for the Wilson coefficients arise as follows:
\begin{equation}
   \underbrace{1 + \alpha_s\ln\frac{M_W^2}{-p^2}}_{\mbox{full theory}}
   = \underbrace{\left( 1 + \alpha_s\ln\frac{M_W^2}{\mu^2}
                 \right)}_{\mbox{$C(\mu)$}}
    \underbrace{\left( 1 + \alpha_s\ln\frac{\mu^2}{-p^2}
                 \right)}_{\mbox{$\langle Q(\mu)\rangle$}} + \dots ,
\end{equation}
where the expression on the left is the matrix element in the full theory 
(which is UV finite and regularized in the IR by an off-shell momentum $p^2$), 
while the expression on the right is the product of a Wilson coefficient and 
a matrix element in the effective theory. The EFT matrix element has the same
dependence on the IR cutoff as the matrix element in the full theory, while
all reference to the fundamental scale $M_W$ resides in the Wilson 
coefficient. Another way of representing this result is in the form of 
logarithmic integrations:
\begin{equation}
   \int_{-p^2}^{M_W^2}\!\frac{dk^2}{k^2}
   = \int_{\mu^2}^{M_W^2}\!\frac{dk^2}{k^2}
   + \int_{-p^2}^{\mu^2}\!\frac{dk^2}{k^2} \,.
\end{equation}
In general, the Wilson coefficients absorb the high-frequency contributions 
of the loop integrals, while the low-frequency contributions reside in the 
EFT matrix elements.
\end{itemize}

\subsection*{FCNC transitions}

As a more interesting example of an effective weak Lagrangian, let us now
consider the important case of flavor-changing neutral current (FCNC) 
transitions \cite{Buchalla:1995vs}. 
Such interactions are absent at tree level in the Standard Model, while they
can be mediated through loop processes. The corresponding suppression of 
FCNC amplitudes provides a window for New Physics searches. The perhaps best
known example is the rare radiative decay $B\to X_s\gamma$, which is
the ``standard candle'' of quark flavor physics.

What are the effective operators that can mediate transitions of the type
$b\to s_L+\mbox{anything}$, where ``anything'' must be flavor diagonal? At
dimension $\delta_i=4$, the two possible operators one can write down (we set 
$m_s=0$ but consider the possibility of having $m_b\ne 0$ for the heavy $b$ 
quark) are $m_b\,\bar s_L b_R$ and $\bar s_L\,i\rlap{\,/}{D}\,b_L$. Note that 
the weak interactions only operate on left-handed fields, so a mass insertion 
${\cal L}_m=m_b(\bar b_R b_L+\bar b_L b_R)$ is needed to turn the left-handed
$b$ quark into a right-handed one. The two operators are equivalent because of
the equation of motion $i\rlap{\,/}{D}\,b_L=m_b\,b_R$. However, the 
``flavor off-diagonal mass term'' $m_b\,\bar s_L b_R$ can always be removed 
by a field redefinition of the quark fields. It follows that there are no
dimension-4 terms in the effective Lagrangian. (Note also that the operator 
$\bar s_L H b_L$ is not gauge invariant.) 

The leading operators in the effective weak Lagrangian once again have 
dimension $\delta_i=6$. They can be arranged in three classes. The first
type of operators is analogous to those in (\ref{example2}):
\begin{eqnarray}\label{4q1}
   && \bar s_L^i\gamma_\mu b_L^i\,\bar q_L^j\gamma^\mu q_L^j \,, \\
   && \bar s_L^i\gamma_\mu b_L^j\,\bar q_L^j\gamma^\mu q_L^i
    \,\,\hat{=}\,\, \bar s_L^i\gamma_\mu q_L^i\,\bar q_L^j\gamma^\mu b_L^j \,. 
    \nonumber
\end{eqnarray}
However, 
since the flavor-singlet quark pair is not restricted to couple to weak 
gauge bosons, we can also have operators of the form
\begin{eqnarray}\label{4q2}
   && \bar s_L^i\gamma_\mu b_L^i\,\bar q_R^j\gamma^\mu q_R^j \,, \\
   && \bar s_L^i\gamma_\mu b_L^j\,\bar q_R^j\gamma^\mu q_R^i
    \,\,\hat{=}\,\, -2 \bar s_L^i q_R^i\,\bar q_R^j b_L^j \,. \nonumber
\end{eqnarray}
Finally, there are operators containing the gauge fields, namely
\begin{equation}
   g_s m_b\,\bar s_L\sigma_{\mu\nu}\,G_a^{\mu\nu} t_a b_R \,, \qquad
   g_s\,\bar s_L\gamma_\nu\,iD_\mu\,G_a^{\mu\nu} t_a b_L \,.
\end{equation}
In the first case a mass insertion is required. The second operator is 
redundant, since the equation of motion for the Yang-Mills field,
\begin{equation}
   D_\mu\,G_a^{\mu\nu} = - \sum_q\,g_s\,\bar q\gamma^\nu t_a q \,,
\end{equation}
implies
\begin{equation}
   g_s\,\bar s_L\gamma_\nu\,iD_\mu\,G_a^{\mu\nu} t_a b_L 
   = - g_s^2 \sum_q\,\bar s_L\gamma_\nu t_q b_L\,\bar q\gamma^\nu t_a q \,,
\end{equation}
which can be written as a linear combination of the four-quark operators in 
(\ref{4q1}) and (\ref{4q2}). When electromagnetic interactions are included, 
we have in addition the operator
\begin{equation}
   e m_b\,\bar s_L\sigma_{\mu\nu}\,F^{\mu\nu} b_R \,,
\end{equation}
where $F^{\mu\nu}$ is the electromagnetic field strength. Also, there are 
additional four-quark operators involving leptons, such as 
\begin{equation}\label{opsll}
   \bar s_L\gamma_\mu b_L\,\bar l_L\gamma^\mu l_L \,, \qquad
   \bar s_L\gamma_\mu b_L\,\bar l_R\gamma^\mu l_R \,, \qquad
   \bar s_L\gamma_\mu b_L\,\bar\nu_{lL}\gamma^\mu\nu_{lL} \,.
\end{equation}

\begin{figure}[t]
\begin{center}
\includegraphics[width=0.64\textwidth]{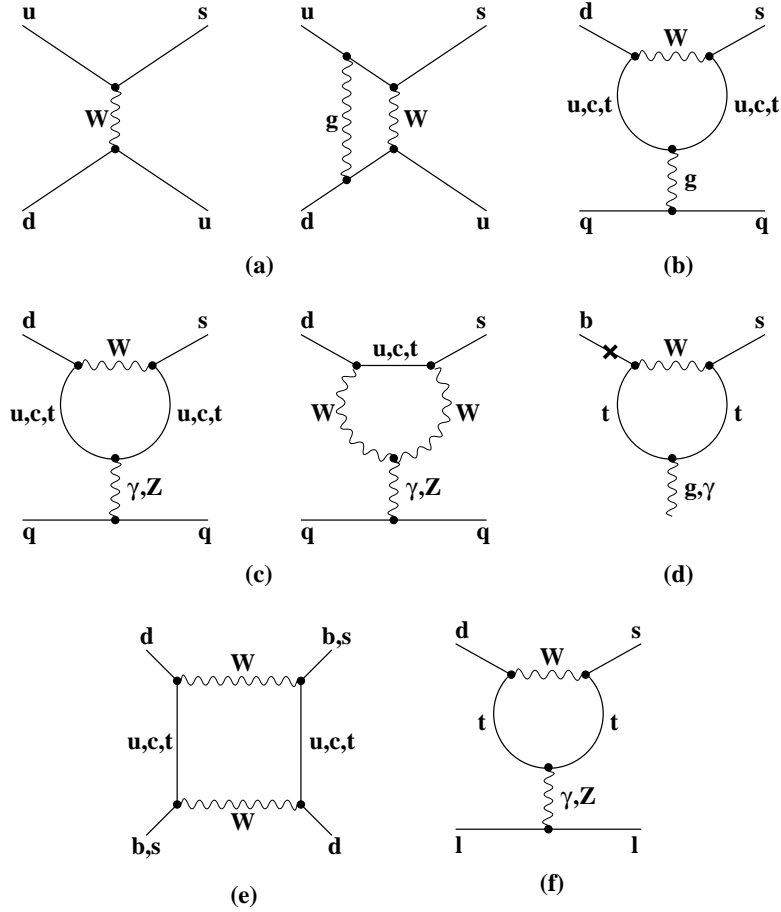} 
\end{center}
\vspace*{-0.5cm}
\caption{\label{fig:graphs}
Typical diagrams in the Standard Model which generate the different operators 
in the effective weak Lagrangian. The current-current operators $Q_{1,2}$ 
result 
from graphs of type~(a), the QCD penguin operators $Q_{3,\dots,6}$ from 
graphs of type~(b), the electroweak penguin operators $Q_{7,\dots,10}$ 
from graphs of type~(c), and the dipole operators from graphs of type~(d). 
Digram~(f) generates the operators with leptons shown in (\ref{opsll}), while 
diagram~(e) contributes to $B$--$\bar B$ and $K$--$\bar K$ mixing.
(Figure taken from \cite{Buchalla:1995vs} with permission from the authors)}
\end{figure}

To summarize, the resulting
Standard Model operator basis for FCNC processes (without leptons, for 
simplicity) contains the ``current-current operators'' (with $p=u,c$)
\begin{eqnarray}
   Q_1^{(p)} &=& (\bar s_i p_i)_{V-A}\,(\bar p_j b_j)_{V-A} \,, \nonumber\\
   Q_2^{(p)} &=& (\bar s_i p_j)_{V-A}\,(\bar p_j b_i)_{V-A} \,, 
\end{eqnarray}
the ``QCD penguin operators''
\begin{eqnarray}
   Q_3 &=& (\bar s_i b_i)_{V-A} \sum_{q=u,d,s,c,b} (\bar q_j q_j)_{V-A}
    \,, \nonumber\\
   Q_4 &=& (\bar s_i b_j)_{V-A} \sum_{q=u,d,s,c,b} (\bar q_j q_i)_{V-A}
    \,, \nonumber\\
   Q_5 &=& (\bar s_i b_i)_{V-A} \sum_{q=u,d,s,c,b} (\bar q_j q_j)_{V+A}
    \,, \nonumber\\
   Q_6 &=& (\bar s_i b_j)_{V-A} \sum_{q=u,d,s,c,b} (\bar q_j q_i)_{V+A} \,,
\end{eqnarray}
the ``electroweak penguin operators'' (with $e_q$ the electric changes of the
quarks in units of $|e|$)
\begin{eqnarray}
   Q_7 &=& (\bar s_i b_i)_{V-A} \sum_{q=u,d,s,c,b} \frac32 e_q\,
    (\bar q_j q_j)_{V+A} \,, \nonumber\\
   Q_8 &=& (\bar s_i b_j)_{V-A} \sum_{q=u,d,s,c,b} \frac32 e_q\,
    (\bar q_j q_i)_{V+A} \,, \nonumber\\
   Q_9 &=& (\bar s_i b_i)_{V-A} \sum_{q=u,d,s,c,b} \frac32 e_q\,
    (\bar q_j q_j)_{V-A} \,, \nonumber\\
   Q_{10} &=& (\bar s_i b_j)_{V-A} \sum_{q=u,d,s,c,b} \frac32 e_q\,
    (\bar q_j q_i)_{V-A} \,,
\end{eqnarray}
and the electromagnetic and chromo-magnetic dipole operators
\begin{eqnarray}
   Q_{7\gamma} &=& - \frac{e m_b}{8\pi^2}\,
    \bar s_L\sigma_{\mu\nu}\,F^{\mu\nu} b_R \,, \nonumber\\
   Q_{8g} &=& - \frac{g_s m_b}{8\pi^2}\,
    \bar s_L\sigma_{\mu\nu}\,G_a^{\mu\nu} t_a b_R \,.
\end{eqnarray}
We use the short-hand notation 
$(\bar q_1 q_2)_{V\pm A}\equiv\bar q_1\gamma^\mu(1\pm\gamma_5) q_2$. Note 
that (if only to confuse you) the standard convention for the electroweak 
penguin operators is such that $Q_{7,8}$ correspond to $Q_{5,6}$, while 
$Q_{9,10}$ correspond to $Q_{3,4}$. Some representative Feynman 
diagrams in the full theory from which these operators originate are shown 
in Figure~\ref{fig:graphs}. The names of the various penguin operators 
reflect the nature of the gauge bosons (gluons for QCD penguins, and 
$\gamma$ or $Z$-bosons for electroweak penguins) emitted from the penguin 
loops.

Some particular features of the Standard Model have been implicitly 
incorporated in the above considerations, namely that only left-handed fields
are involved in flavor-changing weak interactions, that light (approximately 
massless) quarks have identical couplings with respect to the strong 
interactions, and that all up-type ($u,c$) and down-type ($d,s,b$) quark 
fields couple identically to the weak force. 

The unitarity of the CKM matrix implies $\lambda_u+\lambda_c+\lambda_t=0$, 
where $\lambda_p\equiv V_{pb} V_{ps}^*$. We will use this relation to 
eliminate CKM factors involving couplings of the top quark. Note also that in
the limit $m_u=m_c=0$ (which is justified at dimension-6 order) 
the penguin graphs
always involve $\lambda_t=-(\lambda_u+\lambda_c)$. The final result for the
effective weak Lagrangian reads
\begin{equation}\label{LeffW}
   {\cal L}_{\rm eff} = - \frac{G_F}{\sqrt2} \left[
   \sum_{p=u,c}\,\lambda_p \left( C_1 Q_1^{(p)} + C_2 Q_2^{(p)} \right)
   + \sum_{i=3,\dots,10,7\gamma,8g} (\lambda_u+\lambda_c)\,C_i Q_i \right] .
\end{equation}
Note that
\begin{equation}
   \frac{\lambda_u}{\lambda_c} = \frac{ V_{ub} V_{us}^*}{ V_{cb} V_{cs}^*}
   \sim e^{-i\gamma}
\end{equation}
has a non-zero, relative CP-violating phase. This allows for the phenomenon
of CP violation from amplitude interference in FCNC processes -- a phenomenon
that is currently being studied extensively at the $B$-factories 
(see, e.g., \cite{Charles:2004jd}). 

Let me finish this discussion by quoting the matching conditions for the 
various operators at the weak scale $\mu=M_W$. They are 
\cite{Buchalla:1995vs}:
\begin{eqnarray}\label{Cimatching}
   C_1(M_W) &=& 1 - \frac{11}{6}\,\frac{\alpha_s(M_W)}{4\pi} \,, \nonumber\\
   C_2(M_W) &=& \frac{11}{2}\,\frac{\alpha_s(M_W)}{4\pi} \,, \nonumber\\
   C_3(M_W) = C_5(M_W) 
   &=& - \frac16\,\widetilde E_0\bigg(\frac{m_t^2}{M_W^2}\bigg)\,
    \frac{\alpha_s(M_W)}{4\pi} \,, \nonumber\\
   C_4(M_W) = C_6(M_W) 
   &=& \frac12\,\widetilde E_0\bigg(\frac{m_t^2}{M_W^2}\bigg)\,
    \frac{\alpha_s(M_W)}{4\pi} \,, \nonumber\\
   C_7(M_W) &=& f\bigg(\frac{m_t^2}{M_W^2}\bigg)\,
    \frac{\alpha(M_W)}{6\pi} \,, \nonumber\\
   C_9(M_W) &=& \left[ f\bigg(\frac{m_t^2}{M_W^2}\bigg)
    + \frac{1}{\sin^2\theta_W}\,g\bigg(\frac{m_t^2}{M_W^2}\bigg) \right]
    \frac{\alpha(M_W)}{4\pi} \,, \nonumber\\
   C_8(M_W) = C_{10}(M_W) &=& 0 \,,
\end{eqnarray}
with
\begin{eqnarray}
   \widetilde E_0(x) &=& - \frac{7}{12} + O(1/x) \,, \nonumber\\
   f(x) &=& \frac{x}{2} + \frac43\ln x - \frac{125}{36} + O(1/x)
    \,, \nonumber\\
   g(x) &=& - \frac{x}{2} - \frac32\ln x + O(1/x) \,,
\end{eqnarray}
and
\begin{eqnarray}
   C_{7\gamma}(M_W) &=& - \frac13 + O(1/x) \,, \nonumber\\
   C_{8g}(M_W) &=& - \frac18 + O(1/x) \,.
\end{eqnarray}
Note that despite of the fact that there is a heavy top-quark running in 
the penguin loops, the Wilson coefficients exhibit non-decoupling, i.e., 
they do not vanish in the
limit where $m_t\to\infty$. The coefficients of the electroweak penguin
operators are even proportional to $m_t^2$ in this limit. This makes 
electroweak penguin operators relevant for phenomenology, even though the are
suppressed by small electroweak coupling constants.

\section{RG-Improved Perturbation Theory}
\label{lec:4}

There are some important technical aspects which we have ignored in the 
discussion of the previous lecture. Recall the one-loop matching results for
the Wilson coefficients $C_1$ and $C_2$ from (\ref{C1C2}):
\begin{eqnarray}
   C_1(\mu) &=& 1 + \frac{3}{N_c}\,\frac{\alpha_s(\mu)}{4\pi}
    \left( \ln\frac{M_W^2}{\mu^2} - \frac{11}{6} \right) + O(\alpha_s^2) \,,
    \nonumber\\
   C_2(\mu) &=& -3\,\frac{\alpha_s(\mu)}{4\pi}
    \left( \ln\frac{M_W^2}{\mu^2} - \frac{11}{6} \right) + O(\alpha_s^2) \,.
\end{eqnarray}
Ideally, we would like to integrate out {\em all\/} high-frequency modes 
perturbatively and then evaluate the remaining EFT matrix elements 
$\langle Q_i(\mu)\rangle$ at some low scale $\mu\sim\mbox{few GeV}$, below
which perturbation theory becomes untrustworthy. The computation of these 
matrix elements must use a non-perturbative
approach such as lattice QCD, heavy-quark expansions, or chiral perturbation
theory. A glance at the above equations shows a potential problem: the
expansion parameter is not $\frac{\alpha_s}{\pi}\sim 0.1$, but 
$\frac{\alpha_s}{\pi}\ln\frac{M_W^2}{\mu^2}\sim 0.8$. The problem is indeed
generic: in the presence of widely separated scales $M\gg\mu$, 
perturbation theory often involves powers of $\alpha_s\ln\frac{M}{\mu}$ rather
than powers of $\alpha_s$. Such large logarithmic terms must be resummed to 
all orders.

While this problem is particularly acute for almost all practical calculations 
in QCD, it is also relevant to theories with smaller coupling constants. 
For instance, when the gauge couplings of the Standard Model are extrapolated
from low energy up to the GUT scale $M_{\rm GUT}\sim 10^{16}$\,GeV, the 
relevant logarithm is $\ln\frac{M_{\rm GUT}^2}{\mu^2}\approx 65$. Resummation
is essential to control such large logarithms even if the coupling constants
are as small as those for the electro-weak interactions of the Standard Model.

The general solution to the problem of large logarithms is called 
``renormalization-group (RG) improved perturbation theory''. It provides a 
reorganization of perturbation theory in which $\alpha_s\ln\frac{M}{\mu}$ is
treated as an $O(1)$ parameter, while $\alpha_s\ll 1$. Large logarithms are 
resummed to all orders in perturbation theory by solving RG equations. The
nomenclature of RG-improved perturbation theory is as follows: At leading 
order (LO) all terms of the form $(\alpha_s\ln\frac{M}{\mu})^n$ 
with $n=0,\dots,\infty$ are resummed. The result is an $O(1)$ contribution to 
the Wilson coefficient functions. 
At next-to-leading order (NLO), one also
resums terms of the form $\alpha_s(\alpha_s\ln\frac{M}{\mu})^n$, all of 
which count as $O(\alpha_s)$, and so on.
Note that in cases where the term with 
$n=0$ is absent (such as for $C_2$), there may be $O(1)$ effects after 
resummation that not seen at tree level in perturbation theory. This happens 
also for the Wilson coefficients of the QCD penguin operators in the effective 
weak Lagrangian. As shown in (\ref{Cimatching}) the matching conditions for 
the coefficients $C_{2,\dots,6}$ start at $O(\alpha_s)$; yet, after RG 
resummation these coefficients become of $O(1)$ and contribute at the same 
order as the Wilson coefficient $C_1$ of the leading current-current operator.

Before we can perform such resummations, we must study in some more detail
the renormalization of the composite operators in the effective Lagrangian.

\subsection*{Anomalous dimensions}

Consider a complete set (a basis) 
\begin{equation}
   \{ Q_i(\mu) \} \,; \quad i=1,\dots,n \,.
\end{equation}
of operators of dimension $\delta$
allowed by the symmetries (quantum numbers, etc.) of a problem.
Recall that by changing the scale $\mu$ one reshuffles terms from the matrix
elements $\langle Q_i\rangle$ into the coefficients $C_i$, leaving the result
for any observable unchanged, i.e.\
\begin{equation}
   {\cal A} = \sum_{i=1}^n\,C_i(\mu)\,\langle Q_i(\mu)\rangle
   = \sum_{i=1}^n\,C_i(\mu-\delta\mu)\,\langle Q_i(\mu-\delta\mu)\rangle \,.
\end{equation}
The fact that physical observables are scale independent implies that
\begin{equation}\label{RGinv}
   \frac{d}{d\ln\mu} \sum_{i=1}^n\,C_i(\mu)\,\langle Q_i(\mu)\rangle
   = 0 \,.
\end{equation}
Since the operator basis is complete, we can expand the logarithmic 
derivative of the operator matrix elements in terms of the same basis 
operators. We write
\begin{equation}
   \frac{d}{d\ln\mu}\,\langle Q_i(\mu)\rangle
   \equiv - \sum_{j=1}^n\,\gamma_{ij}(\mu)\,\langle Q_j(\mu)\rangle \,.
\end{equation}
If there is more than one operator present, we say that the operators mix 
under scale variation. The dimensionless coefficients $\gamma_{ij}$ measure
the incremental change under scale variation and are free of large
logarithms. They are called anomalous dimensions. Using this definition, 
it follows from (\ref{RGinv}) that
\begin{equation}
   \sum_{j=1}^n \left[ \frac{d}{d\ln\mu}\,C_j(\mu)
   - \sum_{i=1}^n\,C_i(\mu)\,\gamma_{ij}(\mu) \right] 
   \langle Q_j(\mu)\rangle = 0 \,.
\end{equation}
Since by assumption the operators $Q_i$ are linearly independent, we conclude 
that
\begin{equation}
   \frac{d}{d\ln\mu}\,C_j(\mu) - \sum_{i=1}^n\,C_i(\mu)\,\gamma_{ij}(\mu)
   = 0 \,.
\end{equation}
This is the RG equation obeyed by the Wilson coefficient functions. In matrix
notation, we can rewrite this equation as
\begin{equation}
   \frac{d}{d\ln\mu}\,\vec{C}(\mu) = \hat\gamma^T(\mu)\,\vec{C}(\mu) \,.
\end{equation}
The dimensionless anomalous-dimension matrix $\hat\gamma$ depends on the scale
$\mu$ only through the running coupling $\alpha_s(\mu)$. Changing variables
from $\ln\mu$ to $\alpha_s(\mu)$, we find
\begin{equation}\label{CRGE}
   \frac{d}{d\alpha_s(\mu)}\,\vec{C}(\mu)
   = \frac{\hat\gamma^T(\alpha_s(\mu))}{\beta(\alpha_s(\mu))}\,\vec{C}(\mu)
   \,,
\end{equation}
where $\beta=d\alpha_s(\mu)/d\ln\mu$ is the QCD $\beta$-function. The initial 
condition for the solution of the RG equation is set by the values 
$\vec{C}(M_W)$ of the Wilson coefficients at the weak scale. 

Equation (\ref{CRGE}) has the same structure as the Heisenberg equation for
the time dependence of the Hamiltonian in quantum field theory.
The unique solution to this equation is
\begin{equation}
   \vec{C}(\mu) = T_\alpha\exp\Bigg[\,\,
   \int\limits_{\alpha_s(M_W)}^{\alpha_s(\mu)}\!\!\!d\alpha\,
   \frac{\hat\gamma^T(\alpha)}{\beta(\alpha)} \Bigg] \vec{C}(M_W) \,.
\end{equation}
The matrix exponential is defined through its Taylor expansion, and 
the symbol $T_\alpha$ means an ordering such that $\hat\gamma^T(\alpha)$ with
larger $\alpha$ stands to the left of those with smaller $\alpha$. Such an
ordering prescription is necessary because, in general, the matrices 
$\hat\gamma^T(\alpha)$ at different $\alpha$ values do not commute.

We now perform a (controlled) perturbative expansion of the quantities
$\vec{C}(M_W)$, $\hat\gamma(\alpha)$, and $\beta(\alpha)$ entering the general
solution, all of which are free of large logarithms. Consider, for simplicity,
the case of a single Wilson coefficient ($n=1$, no mixing). Writing
\begin{equation}
   \gamma(\alpha_s) = \gamma_0\,\frac{\alpha_s}{4\pi} + O(\alpha_s^2)
    \,, \quad
   \beta(\alpha_s) = -2\alpha_s \left[ \beta_0\,\frac{\alpha_s}{4\pi} + 
    O(\alpha_s^2) \right] , \quad
   C(M_W) = 1 + O(\alpha_s) \,,
\end{equation}
we find the leading-order solution
\begin{equation}
   C(\mu) = \left( \frac{\alpha_s(\mu)}{\alpha_s(M_W)}
   \right)^{-\frac{\gamma_0}{2\beta_0}} \Big[ 1 + O(\alpha_s) \Big] \,.
\end{equation}
To see that this sums the leading logarithms, note that
\begin{equation}
   \left( \frac{\alpha_s(\mu)}{\alpha_s(M_W)}
   \right)^{-\frac{\gamma_0}{2\beta_0}}
   \approx \left( 1 + \beta_0\,\frac{\alpha_s}{4\pi}\,\ln\frac{M_W^2}{\mu^2}
   \right)^{-\frac{\gamma_0}{2\beta_0}}
   = 1 - \frac{\gamma_0}{2}\,\frac{\alpha_s}{4\pi}\,\ln\frac{M_W^2}{\mu^2}
   + O(\alpha_s^2\ln^2\frac{M_W^2}{\mu^2}) \,.
\end{equation}

It is straightforward to go to higher orders in the expansion in $\alpha_s$. 
For the case of a single operator, the NLO solution reads
\begin{equation}
   C(\mu) = \left( \frac{\alpha_s(\mu)}{\alpha_s(M_W)}
   \right)^{-\frac{\gamma_0}{2\beta_0}} \left[ 1
   + \frac{\alpha_s(\mu)-\alpha_s(M_W)}{4\pi}\,S
   + c_1\,\frac{\alpha_s(M_W)}{4\pi} + O(\alpha_s^2) \right] ,
\end{equation}
where
\begin{equation}
   S = - \frac{\gamma_0}{2\beta_0} \left( \frac{\gamma_1}{\gamma_0}
   - \frac{\beta_1}{\beta_0} \right) ,
\end{equation}
and we have expanded
\begin{eqnarray}
   \gamma(\alpha_s) &=& \sum_{n=0}^\infty \gamma_n
    \left( \frac{\alpha_s}{4\pi} \right)^{n+1} , \qquad
   \beta(\alpha_s) = -2\alpha_s \sum_{n=0}^\infty \beta_n
    \left( \frac{\alpha_s}{4\pi} \right)^{n+1} , \nonumber\\
   C(M_W) &=& 1 + \sum_{n=1}^\infty c_n
    \left( \frac{\alpha_s(M_W)}{4\pi} \right)^n .
\end{eqnarray}
The generalization to the case of operator mixing is discussed in the 
literature \cite{Buchalla:1995vs}.

The systematics of RG-improved perturbation theory is summarized in the 
following table:
\begin{center}
\begin{tabular}{c|ccc}
Order & $\gamma$, $\beta$ & $C(M_W)$ \\
\hline
LO & 1-loop & tree-level \\
NLO & 2-loop & 1-loop \\
NNLO & 3-loop & 2-loop
\end{tabular}
\end{center}
The LO approximation is really only good for illustration purposes. At NLO we
achieve the same accuracy as in the case of a conventional one-loop 
calculation for a single-scale problem. Note, however, that two-loop anomalous 
dimensions are required at this order. The NNLO
approximation is the state of the art for many applications, where high 
precision is of concern.

\section{Effective Theories for Heavy Quarks}

Heavy-quark systems provide prime examples for applications of the EFT 
technology, because the hierarchy $m_b\gg\Lambda_{\rm QCD}$ provides a natural
separation of scales. Physics at the scale $m_b$ is of a short-distance 
nature, while for heavy-quark systems there is always also some hadronic 
physics governed by the confinement scale $\Lambda_{\rm QCD}$. Being able to 
separate the short-distance and long-distance effects associated with these
two scales is vital for any quantitative description in heavy-quark physics.
For instance, if the long-distance hadronic matrix elements are obtained from
lattice QCD, then it is necessary to analytically 
compute the short-distance effects, which
come from short-wavelength modes that do not fit on present-day lattices. 
In many other instances, the long-distance hadronic physics can
be encoded in a small number of universal parameters. To
identify these parameters requires that one first extracts all short-distance 
effects.

\subsection*{Heavy-quark effective theory (HQET)}

The simplest effective theory for heavy-quark systems is heavy-quark 
effective theory (HQET) \cite{Georgi:1990um}. It provides a simplified 
description of the soft interactions 
of a single heavy quark interacting with soft, light partons. This includes the
interactions that bind the heavy quark with other light partons inside heavy
mesons ($B$, $B^*$, \dots) and baryons ($\Lambda_b$, $\Sigma_b$, \dots). I will
only offer an HQET primer in this lecture, referring to reviews in the 
literature for a more detailed discussion 
\cite{Neubert:1993mb,Manohar:2000dt}. 

A softly interacting heavy quark is nearly on-shell. Its momentum may be 
decomposed as
\begin{equation}
   p_Q^\mu = m_Q\,v^\mu + k^\mu \,,
\end{equation}
where $v$ is the 4-velocity of the hadron containing the heavy quark 
($v^2=1$), and the ``residual momentum'' $k\sim\Lambda_{\rm QCD}$. This 
off-shell momentum results from the soft interactions of the heavy quark with 
its environment. (This 
assumes a reference frame in which the heavy meson has a small velocity, 
$v=O(1)$.) A near on-shell Dirac spinor has two large and two small 
components. We define
\begin{equation}\label{hdef}
   Q(x) = e^{-im_Q v\cdot x} \left[ h_v(x) + H_v(x) \right] ,
\end{equation}
where
\begin{equation}
   h_v(x) = e^{im_Q v\cdot x}\,\frac{1+\rlap/v}{2}\,Q(x)
\end{equation}
are the large, ``upper'' components, while
\begin{equation}
   H_v(x) = e^{im_Q v\cdot x}\,\frac{1-\rlap/v}{2}\,Q(x)
\end{equation}
are the small, ``lower'' components. The extraction of the phase factor in 
(\ref{hdef}) implies that the fields $h_v$ and $H_v$ carry the residual 
momentum $k$. These fields obey the projection relations
\begin{equation}\label{project}
   \rlap/v\,h_v(x) = h_v(x) \,, \qquad
   \rlap/v\,H_v(x) = - H_v(x) \,.
\end{equation}
Inserting these definitions into the Dirac Lagrangian yields
\begin{eqnarray}\label{LQ}
   {\cal L}_Q &=& \bar Q\,(i\rlap{\,/}{D} - m_Q)\,Q \nonumber\\
   &=& \bar h_v\,i\rlap{\,/}{D}\,h_v
    + \bar H_v\,(i\rlap{\,/}{D} - 2m_Q)\,H_v
    + \bar h_v\,i\rlap{\,/}{D}\,H_v
    + \bar H_v\,i\rlap{\,/}{D}\,h_v \nonumber\\
   &=& \bar h_v\,iv\cdot D\,h_v
    + \bar H_v\,(-iv\cdot D-2m_Q)\,H_v
    + \bar h_v\,i\vec{\rlap{\,/}{D}}\,H_v
    + \bar H_v\,i\vec{\rlap{\,/}{D}}\,h_v \,,
\end{eqnarray}
where $i\vec{D}^\mu=iD^\mu-v^\mu\,iv\cdot D$ is the ``spatial'' covariant 
derivative (note that $v^\mu=(1,\vec{0})$ in the heavy-hadron rest frame).
We have used that between two $h_v$ spinors a Dirac matrix $\gamma^\mu$ can 
be replaced with $v^\mu$, while between two $H_v$ spinors it can 
be replaced with $-v^\mu$. We have also used the projection properties 
(\ref{project}), which in particular imply that $\bar H_v\,\rlap/v\,h_v=0$.

The interpretation of (\ref{LQ}) is that the field $h_v$ describes a massless
fermion, while $H_v$ describes a heavy fermion with mass $2m_Q$. Both modes
are coupled to each other via the last two terms. Soft interactions cannot
excite the heavy fermion, so we integrate it out from the generating functional
of the theory. The light field which remains describes the fluctuations of
the heavy quark about its mass shell. Solving the classical equation of motion 
for the field $H_v$ yields
\begin{equation}\label{Hv}
   H_v = \frac{1}{2m_Q+iv\cdot D}\,i\vec{\rlap{\,/}{D}}\,h_v
   = \frac{1}{2m_Q} \sum_{n=0}^\infty \left( - \frac{iv\cdot D}{2m_Q}
   \right)^n i\vec{\rlap{\,/}{D}}\,h_v \,,
\end{equation}
which implies $H_v=O(\Lambda_{\rm QCD}/m_Q)\,h_v$ provided the
residual momenta are small. The leading-order effective Lagrangian obtained 
from (\ref{LQ}) then reads
\begin{equation}
   {\cal L}_{\rm HQET} = \bar h_v\,iv\cdot D_s\,h_v + O(1/m_Q) \,.
\end{equation}
Note that the covariant derivative 
$iD_s^\mu=i\partial^\mu+g_s A_s^\mu$ contains
only the soft gluon field. Hard gluons have been integrated out. 

It is straightforward to include power corrections to the effective 
Lagrangian by keeping higher-order terms in (\ref{Hv}). One
finds that at subleading order in $1/m_Q$ two new operators arise, such that
\begin{equation}\label{LHQET2}
   {\cal L}_{\rm HQET} = \bar h_v\,iv\cdot D_s\,h_v 
   + \frac{1}{2m_Q} \left[ \bar h_v\,(i\vec{D}_s)^2 h_v
   + C_{\rm mag}(\mu)\,\frac{g_s}{2}\,\bar h_v\,\sigma_{\mu\nu}\,
   G_s^{\mu\nu} h_v \right] + \dots \,.
\end{equation}
The new operators are referred to ask the ``kinetic energy'' and the 
``chromo-magnetic interaction''. The kinetic-energy operator corresponds to 
the first correction term in the Taylor expansion of the relativistic
energy $E=\sqrt{\vec{p}^{\,\,2}+m_Q^2}=m_Q+\vec{p}^{\,\,2}/2m_Q+\dots$, and 
Lorentz invariance ensures that its coefficient is not renormalized. The 
Wilson coefficient of the chromo-magnetic interaction operators is, however, 
non-trivial. It has been calculated in \cite{Amoros:1997rx} at NLO in 
RG-improved perturbation theory.

The leading term in the HQET Lagrangian exhibits a SU(2$n_Q$) spin-flavor 
symmetry. Its physical meaning is that, in the infinite mass limit, 
the properties of hadronic systems containing a single heavy quark are 
insensitive to the spin and flavor of the heavy quark \cite{Isgur:1989vq}. 
The flavor symmetry is broken by the operators arising at order $1/m_Q$ and 
higher. Note, however, that at this order only the chromo-magnetic operator
breaks the spin symmetry. Many phenomenological implications of heavy-quark
symmetry are explored in \cite{Neubert:1993mb}.

\subsubsection*{Scalings of fields}

It is useful to set up a systematic power counting in 
$\lambda=\Lambda_{\rm QCD}/m_Q$ by assigning scaling properties for all 
objects in the effective theory. To do this, all dimensionful quantities
are expressed in powers of the fundamental scale $m_Q$. For the residual
momentum we find $k^\mu\sim\Lambda_{\rm QCD}=\lambda m_Q\sim\lambda$. 
All derivatives on the 
soft fields in HQET obey the same scaling as the residual momentum, i.e.\
$\partial^\mu\sim\lambda$.

The scaling property of the heavy-quark field $h_v$ follows by considering
the free quark propagator in position space. We have
\begin{equation}
   \langle 0|\,T\,\{ h_v(x)\,\bar h_v(0) \}\,|0\rangle
   = \int \frac{d^4k}{(2\pi)^4}\,e^{-ik\cdot x}\,\frac{i}{v\cdot k+i\epsilon}
   \sim \lambda^4\cdot\frac{1}{\lambda}\sim \lambda^3 \,,
\end{equation}
from which it follows that $h_v\sim\lambda^{3/2}$. For the soft gluon field,
a similar consideration shows that
\begin{equation}
   \langle 0|\,T\,\{ A_s^\mu(x)\,A_s^\nu(0) \}\,|0\rangle
   = \int \frac{d^4k}{(2\pi)^4}\,e^{-ik\cdot x}\,\frac{i}{k^2+i\epsilon}
   \left[ - g^{\mu\nu} + (1-a)\,\frac{k^\mu k^\nu}{k^2} \right]
   \sim \lambda^4\cdot\frac{1}{\lambda^2}\sim \lambda^2 \,,
\end{equation}
and hence $A_s^\mu\sim\lambda$. Note that this ensures a homogeneous scaling
of the covariant derivative, $D_s^\mu\sim\lambda$.
Using these relations, we see that the different terms in the HQET Lagrangian
in (\ref{LHQET2}) scale like
\begin{equation}
   {\cal L}_{\rm HQET} \sim \lambda^4 + \frac{\lambda^5}{m_Q} + \dots \,.
\end{equation}
The scaling of the integration measure $d^4x$ for operators containing soft 
fields follows from the requirement that $e^{-ik\cdot x}\sim O(1)$, since this
defines what are ``important contributions'' to Fourier integrals. We conclude 
that $x^\mu\sim\lambda^{-1}$, $d^4x\sim\lambda^{-4}$, and hence
\begin{equation}
   S_{\rm HQET} = \int d^4x\,{\cal L}_{\rm HQET}(x)
   \sim \lambda^0 + \frac{\lambda}{m_Q} + \dots \,,
\end{equation}
as it should be.

\subsubsection*{Residual gauge invariance}

The effective Lagrangian ${\cal L}_{\rm HQET}$ defines an effective theory
for soft interactions of heavy quarks. All hard interactions (including the
couplings of heavy quarks to hard gluons) are integrated out in the 
construction of the effective Lagrangian. As a result, the effective
theory no longer has the full gauge invariance of QCD, but only a residual
gauge invariance with respect to gauge transformations that preserve the 
scaling properties of the fields. These are called ``soft gauge 
transformations'' and denoted by $U_s(x)$. The transformation rules are
\begin{eqnarray}\label{residualgauge}
   h_v(x) &\to& U_s(x)\,h_v(x) \,, \nonumber\\
   A_s^\mu(x) &\to& U_s(x)\,A_s^\mu(x)\,U_s^\dagger(x)
    + \frac{i}{g_s}\,U_s(x)\,[\partial^\mu,U_s^\dagger(x)] \,. 
\end{eqnarray}
Operationally, ``soft'' functions like $A_s^\mu(x)$ and $U_s(x)$ can be defined
via a restriction to soft modes in their Fourier decomposition. In practice, 
however, the use of dimensional regularization makes it unnecessary to 
introduce the hard cutoffs associated with this construction.

\subsubsection*{Decoupling transformation}

The couplings of soft gluons to heavy quarks can be ``removed'' by the
(non-local, field-dependent) field redefinition
\begin{equation}\label{hvdecoupl}
   h_v(x) = S_v(x)\,h_v^{(0)}(x) \,,
\end{equation}
with
\begin{equation}\label{Sdef}
   S_v(x) = P\exp\left( i g_s \int_{-\infty}^0\!dt\,v\cdot A_s(x+tv) \right)
\end{equation}
a time-like Wilson line extending from minus infinity to the point $x$. The 
symbol $P$ means an ordering with respect to $t$ such that gauge fields are 
ordered from left to right in the order of decreasing $t$ values. The Wilson 
line $S_v^\dagger(x)$ is given by a similar expression but with the opposite
ordering prescription, and with $ig_s$ replaced by $-ig_s$ in the exponent.

The soft Wilson line obeys the important property
\begin{equation}\label{Srule}
   S_v^\dagger\,iv\cdot D_s\,S_v = iv\cdot\partial \,.
\end{equation}
Using this relation, it follows that in terms of the new fields the HQET 
Lagrangian becomes
\begin{equation}
   {\cal L}_{\rm HQET} = \bar h_v^{(0)}\,iv\cdot\partial\,h_v^{(0)}
   + O(1/m_Q) \,.
\end{equation}
At leading order in $1/m_Q$, this is a free theory as far as the 
strong interactions of heavy quarks are concerned! However, the theory is 
nevertheless non-trivial once we allow for the presence of 
external sources. Consider, e.g., what happens to a flavor-changing 
weak-interaction current, which turns a heavy $b$-quark into a $c$-quark
(plus a $W$ boson). At tree level, matching such a current onto HQET gives
\begin{equation}\label{newcurrent}
   \bar c\gamma^\mu(1-\gamma_5) b
   \to \bar h_{v'}\gamma^\mu(1-\gamma_5) h_v
   = \bar h_{v'}^{(0)}\gamma^\mu(1-\gamma_5)
    (S_{v'}^\dagger S_v) h_v^{(0)} \,.
\end{equation}
Here $v$ and $v'$ are the (in general different) velocities of the heavy 
mesons containing the heavy quarks. Unless the
two velocities are equal, the object $S_{v'}^\dagger S_v$ is non-trivial, and 
hence the soft gluons do not decouple from the heavy quarks inside the 
current operator. Indeed, if we close the integration contour at $t=-\infty$
we may interpret $S_{v'}^\dagger S_v$ as a Wilson loop with a cusp at the 
point $x$, where the two paths parallel to the different velocity vectors 
intersect. The presence of the cusp leads to non-trivial UV behavior 
(for $v\ne v'$), which is described by a cusp anomalous dimension 
$\Gamma_c(w)$ depending on the kinematical invariant $w=v\cdot v'$. This 
quantity was calculated at two-loop order as early as in 1987 
\cite{Korchemsky:1987wg}. The cusp anomalous dimension is nothing but the 
celebrated velocity-dependent anomalous dimension of heavy-quark currents, 
which was rediscovered three years later in the context of HQET 
\cite{Falk:1990yz}.

The technology introduced in the last three subsections is usually not taught 
in courses on HQET, and indeed it is not required necessarily. However, the 
interpretation of heavy quarks as Wilson lines is very useful, and it was 
put forward in some of the very first papers on the subject 
\cite{Eichten:1989zv}. I have emphasized this technology here, 
because it will be useful in the study of the interactions of heavy 
quarks with collinear degrees of freedom, to which we turn now.

\subsection*{Soft-collinear effective theory (SCET)}

An long-standing problem in QCD is how to systematically parameterize 
long-distance effects (power corrections) for processes that do not
admit a local OPE. In problems with a large mass 
$M$ or a large Euclidean momentum transfer $Q^2$, the OPE provides a rigorous
framework for an expansion of matrix elements in powers and logarithms of 
the large scale. However, processes involving energetic light particles
pose new challenges. Here some components of $p^\mu$ are large, but 
$p^2\approx 0$ is small. The kinematics in these ``jet-like'' processes is 
intrinsically Minkowskian, and the separation of short-distance from 
long-distance physics becomes a tricky issue. In particular, lattice QCD cannot
be used (not even in principle) to study non-perturbative physics near the 
light cone.

There are many important examples of physical systems where energetic light
partons play an important role. They include jet physics (e.g., the 
determination 
of $\alpha_s$), rare exclusive $B$-meson decays such as $B\to\pi\pi$ and 
$B\to\phi K_S$ (unitarity-triangle physics, physics beyond the Standard 
Model), inclusive $B$ decays such as $B\to X_s\gamma$, and many more. 
For concreteness, consider the two examples $B\to X_s\gamma$ and $B\to\pi\pi$. 
In the first case, working in the rest frame of the initial $B$ meson and 
choosing the $z$-direction as the direction of the jet $X_s$, the 
momenta of the outgoing particles can be written (now in ordinary 4-vector 
notation) as 
\begin{equation}
  p_X^\mu = (M_B-E_\gamma,0,0,E_\gamma) \,, \qquad
  p_\gamma^\mu = (E_\gamma,0,0,-E_\gamma) \,.
\end{equation}
Experimentally one is forced to cut on 
large photon energy, $E_\gamma\approx M_B/2$, such that 
$M_B-2E_\gamma=O(\Lambda_{\rm QCD})$. This implies large energy 
$E_X\approx M_B/2$ and small invariant mass 
$M_X^2=M_B(M_B-2E_\gamma)=O(M_B\Lambda_{\rm QCD})$ for the hadronic final 
state. The example $B\to\pi\pi$ is even simpler. In this case, we have
\begin{equation}
  p_1^\mu = (E_\pi,0,0,\sqrt{E_\pi^2-m_\pi^2}) \,, \qquad
  p_2^\mu = (E_\pi,0,0,-\sqrt{E_\pi^2-m_\pi^2}) \,,
\end{equation}
where $E_\pi=M_B/2$. The final-state pions are on-shell, 
$p_1^2=p_2^2=m_\pi^2$.

The question is: What is there to integrate out in these processes? The 
large scales are not just given by the masses of heavy 
particles, but also by the large energies of some fast light particles. 
Nevertheless, the presence of several different scales means that 
we can classify quantum fluctuations as hard, hard-collinear, collinear, or
soft. For the two examples mentioned above, the corresponding scales are
\begin{eqnarray}
   \mbox{hard:} \quad && M_B,~ E_\gamma \nonumber\\
   \mbox{hard-collinear:} \quad && M_X^2\sim M_B\Lambda_{\rm QCD}
    \nonumber\\
   \mbox{soft:} \quad && \Lambda_{\rm QCD}
\end{eqnarray}
in the case of $B\to X_s\gamma$, and 
\begin{eqnarray}
   \mbox{hard:} \quad && M_B,~ E_\pi \nonumber\\
   \mbox{collinear:} \quad && m_\pi \nonumber\\
   \mbox{soft:} \quad && \Lambda_{\rm QCD}
\end{eqnarray}
in the case of $B\to\pi\pi$.

Our first goal is to integrate out all hard quantum fluctuations. 
For processes with light-like kinematics
this will lead to a {\em non-local\/} EFT \cite{Bauer:2000yr,Bauer:2001ct,%
Bauer:2001yt,Chay:2002vy,Beneke:2002ph,Hill:2002vw}, which will provide us 
with a systematic generalization of the concept of a local OPE. Moreover, 
because there are several short-distance scales in the problem (hard and 
hard-collinear), the construction of the EFT often proceeds in several steps.
The last subsection in this lecture provides an explicit example.

\subsubsection*{Power counting}

The expansion parameter of SCET is $\lambda=\Lambda_{\rm QCD}/E$, where 
$E$ is the typical energy of the (hard-) collinear particles. In $B$ decays, 
$E\sim m_b$, but it is useful to keep these two parameters separate for the 
time being. Because of the presence of fast, collinear particles it is 
convenient to decompose 4-vectors in a light-cone basis spanned by two 
light-like vectors $n$ and $\bar n$ and two transverse coordinates. 
We have $n^2=\bar n^2=0$ and choose $n\cdot\bar n=2$. If the collinear 
particles are aligned along the $z$ direction, we choose $n^\mu=(1,0,0,1)$
and $\bar n^\mu=(1,0,0,-1)$. In other words, $n$ points in the direction of
the collinear momenta, and $\bar n$ points in the opposite direction. For an
arbitrary 4-vector $p^\mu$, we write
\begin{eqnarray}
   p^\mu &=& \frac{\bar n^\mu}{2}\,n\cdot p + \frac{n^\mu}{2}\,\bar n\cdot p
    + p_\perp^\mu \nonumber\\
   &\equiv& p_+^\mu + p_-^\mu + p_\perp^\mu
    \equiv (p_+,p_-,p_\perp) \,,
\end{eqnarray}
where I have introduced several useful notations in the second line. Let us 
perform the light-cone decomposition for the relevant momenta in the two
examples discussed above. For $B\to X_s\gamma$ decay, the momenta of the
final-state particles are decomposed as
\begin{equation}
  p_X^\mu = M_B\,\frac{n^\mu}{2} + (M_B-2E_\gamma)\,\frac{\bar n^\mu}{2} \,,
   \qquad
  p_\gamma^\mu = 2E_\gamma\,\frac{\bar n^\mu}{2} \,.
\end{equation}
For the hadronic final-state jet we have $\bar n\cdot P_X=M_B$ and 
$n\cdot P_X=M_B-2E_\gamma\sim\Lambda_{\rm QCD}$. For the example of 
$B\to\pi\pi$ decay, we have 
\begin{eqnarray}
  p_1^\mu
  &=& \frac{M_B}{2}\,\bigg( 1 + \sqrt{1-\frac{4m_\pi^2}{M_B^2}} \bigg)\,
   \frac{\bar n^\mu}{2} 
   + \frac{M_B}{2}\,\bigg( 1 - \sqrt{1-\frac{4m_\pi^2}{M_B^2}} \bigg)\,
   \frac{\bar n^\mu}{2} \,, \nonumber\\
  p_2^\mu
  &=& \frac{M_B}{2}\,\bigg( 1 - \sqrt{1-\frac{4m_\pi^2}{M_B^2}} \bigg)\,
   \frac{\bar n^\mu}{2} 
   + \frac{M_B}{2}\,\bigg( 1 + \sqrt{1-\frac{4m_\pi^2}{M_B^2}} \bigg)\,
   \frac{\bar n^\mu}{2} \,,
\end{eqnarray}
and hence $\bar n\cdot p_1=n\cdot p_2\approx M_B$ and 
$n\cdot p_2=\bar n\cdot p_1\approx m_\pi^2/M_B$.

We now distinguish between
different types of momenta, classified according to their scaling properties
with the large energy $E\gg\Lambda_{\rm QCD}$. For the example of 
$B\to X_s\gamma$ decay, the relevant momenta are 
(with $\Lambda\sim\Lambda_{\rm QCD}$):\footnote{Unfortunately, the 
terminology for the different momentum modes in SCET is not unique. Some 
authors refer to hard-collinear modes as collinear ones, and to soft modes 
as ultra-soft ones.} 
\begin{eqnarray}\label{SCET1scale}
   \mbox{hard:} \quad && p_h^\mu\sim (E,E,E)\sim (1,1,1) \nonumber\\
   \mbox{hard-collinear:} \quad
   && p_{hc}^\mu\sim (\Lambda,E,\sqrt{E\Lambda})
    \sim (\lambda,1,\lambda^{1/2}) \nonumber\\
   \mbox{soft:} \quad && p_s^\mu\sim (\Lambda,\Lambda,\Lambda)
    \sim (\lambda,\lambda,\lambda)
\end{eqnarray}
The corresponding virtualities are $p_h^2\sim E^2$, $p_{hc}^2\sim E\Lambda$, 
and $p_s^2\sim\Lambda^2$. The partons of the final-state hadronic jet 
typically carry hard-collinear momenta, however, the jet may also contain some 
partons with soft momenta.
The effective theory results from integrating out all fields carrying hard
momenta, keeping fields with hard-collinear or soft momenta as dynamical 
degrees of freedom.

In exclusive processes such as $B\to\pi\pi$ one also encounters particles 
with collinear momenta, whose virtualities are $p_c^2\sim\Lambda^2$. The 
corresponding scaling relations are:
\begin{eqnarray}
   \mbox{collinear-1:} \quad && p_{c1}^\mu\sim (\Lambda^2/E,E,\Lambda)
    \sim (\lambda^2,1,\lambda) \nonumber\\
   \mbox{collinear-2:} \quad && p_{c2}^\mu\sim (E,\Lambda^2/E,\Lambda)
    \sim (1,\lambda^2,\lambda)
\end{eqnarray}
While this looks similar to the scaling relation for a hard-collinear 
momentum, just with $\lambda$ replaced by $\lambda^2$, the difference is that 
the same replacement is {\em not\/} done for the soft fields. Hence, in a 
situation where both soft and collinear fields are present, integrating out
hard modes results in a different EFT than the one mentioned above. It
is conventional to call the EFT for soft and hard-collinear fields SCET-1, and
that for soft and collinear fields SCET-2. In the latter case, the theory 
also contains so-called messenger modes with momentum scaling 
$p_m^\mu\sim(\lambda^2,\lambda,\lambda^{3/2})$. We will not discuss the 
many subtleties of SCET-2 here but refer the interested reader to the 
literature \cite{Becher:2003qh,Becher:2003kh}. From now on we study the theory 
SCET-1 in further detail, calling it SCET for simplicity.

As mentioned above, after integrating out hard modes (i.e., modes carrying 
hard momenta) we obtain an EFT for soft and hard-collinear fields. The 
momentum scalings in (\ref{SCET1scale}) allow for interactions among these 
modes subject to the constraint that 
interactions among hard-collinear and soft fields are only allowed if the soft
fields couple to at least two hard-collinear fields. The reason is that, 
when we couple one or more soft fields to a 
hard-collinear field, the resulting total momentum scales like a 
hard-collinear momentum: 
$(\lambda,1,\lambda^{1/2})+(\lambda,\lambda,\lambda)%
\sim(\lambda,1,\lambda^{1/2})$. Momentum conservation then implies that there 
is another hard-collinear particle that can absorb this momentum.

Let us now introduce the various fields of SCET obtained by decomposing 
the quark and gluon fields into various momentum modes. The soft fields and
their scalings are the same as in HQET:
\begin{eqnarray}
   \mbox{soft heavy quark:} \quad && h_v\sim\lambda^{3/2} \nonumber\\
   \mbox{soft light quark:} \quad && q_s\sim\lambda^{3/2} \nonumber\\
   \mbox{soft gluon:} \quad && A_s^\mu\sim\lambda
\end{eqnarray}
Next, we introduce the following hard-collinear fields:
\begin{eqnarray}
   \mbox{hard-collinear light quark:} \quad && \xi\sim\lambda^{1/2}
    \nonumber\\
   \mbox{hard-collinear gluon:} \quad
   && A_{hc}^\mu\sim(\lambda,1,\lambda^{1/2})
\end{eqnarray}
There are no hard-collinear heavy-quark fields, since the corresponding 
virtualities would be hard, $p_b^2-m_b^2=(m_b\,v+k_{hc})^2-m_b^2%
\approx 2m_b\,v\cdot k_{hc}\approx 2m_b E_{hc}$, and so these fields have 
already been removed from the effective theory. Note that the
scaling of the hard-collinear gluons is such that the covariant derivative
$iD_{hc}^\mu=i\partial^\mu+g_s A_{hc}^\mu$ has a homogeneous scaling when
acting on hard-collinear fields.

As in the case of HQET, the scaling properties of the hard-collinear 
fields can be derived most easily by analyzing the corresponding two-point 
functions \cite{Beneke:2002ph}. For the gluon propagator, we have
\begin{equation}
   \langle 0|\,T\{ A_{hc}^\mu(x)\,A_{hc}^\nu(y) \}\,|0\rangle
   = \int\frac{d^4p}{(2\pi)^4}\,e^{-ip\cdot x}\,\frac{i}{p^2+i\epsilon}
   \left[ - g^{\mu\nu} + (1-a)\,\frac{p^\mu p^\nu}{p^2} \right] .
\end{equation}
Generically (i.e., for $a\ne 1$) this scales like $p^\mu p^\nu$, and
hence in an arbitrary gauge $A_{hc}^\mu$ scales like a hard-collinear momentum
$p^\mu$. Note that it would be a mistake to
assign the scaling $A_{hc}^\mu\sim\lambda^{1/2}$ to the gluon field that one 
finds in Feynman 
gauge ($a=1$), even if Feynman gauge is adopted in practical calculations.
(This mistake has been made in the literature.)

Let us now discuss the hard-collinear fermion field. For simplicity we will 
assume zero mass from now on, even though it would be straightforward to 
include a mass term of $O(\lambda^{1/2})$ or smaller in the effective 
Lagrangian. We decompose the 4-component Dirac spinor field into two fields
\begin{equation}
   \psi_{hc}(x) = \xi(x) + \eta(x) \,, \qquad
   \mbox{with} \quad
   \rlap/n\,\xi = 0 \,, \quad \rlap/\bar n\,\eta = 0 \,.
\end{equation}
Using that
\begin{equation}
   \frac{\rlap/n\rlap/\bar n}{4} + \frac{\rlap/\bar n\rlap/n}{4}
   = \frac{2n\cdot\bar n}{4} = 1 \,,
\end{equation}
it is easy to see that $\frac14\,\rlap/n\rlap/\bar n$ and 
$\frac14\,\rlap/\bar n\rlap/n$ are projection operators onto these two 
2-component spinors, such that
\begin{equation}
   \xi = \frac{\rlap/n\rlap/\bar n}{4}\,\psi_{hc} \,, \qquad
   \eta = \frac{\rlap/\bar n\rlap/n}{4}\,\psi_{hc} \,.
\end{equation}
Consider now the two-point function
\begin{equation}
   \langle 0|\,T\{ \psi_{hc}(x)\,\bar\psi_{hc}(y) \}\,|0\rangle
   = \int\frac{d^4p}{(2\pi)^4}\,e^{-ip\cdot x}\,\frac{i}{p^2+i\epsilon}\,
   (\rlap/p_+ + \rlap/p_- + \rlap/p_\perp) \,.
\end{equation}
Projecting onto the various components, we find
\begin{eqnarray}
   && \langle 0|\,T\{ \xi(x)\,\bar\xi(y) \}\,|0\rangle\sim\lambda \,, 
    \nonumber\\
   && \langle 0|\,T\{ \eta(x)\,\bar\eta(y) \}\,|0\rangle\sim\lambda^2 \,,
    \nonumber\\
   && \langle 0|\,T\{ \xi(x)\,\bar\eta(y) \}\,|0\rangle \,, ~ 
    \langle 0|\,T\{ \eta(x)\,\bar\xi(y) \}\,|0\rangle\sim\lambda^{3/2} \,.
\end{eqnarray}
Obviously, this implies $\xi\sim\lambda^{1/2}$ and $\eta\sim\lambda$. It
follows that a hard-collinear fermion spinor has two ``large'' and two 
``small'' components. The small-component field $\eta$ can be integrated out
from the theory, leaving $\xi$ as the field describing hard-collinear fermions 
in SCET. Like the soft heavy quarks in HQET, hard-collinear quarks in SCET
are thus described by effective 2-component fields.

\subsubsection*{Effective Lagrangian}

Let us first derive how hard-collinear quarks interact with hard-collinear or 
soft gluons, following closely the derivation presented in 
\cite{Beneke:2002ph} (see also \cite{Bauer:2000yr,Bauer:2001yt}). When 
expressed in terms of $\xi$ and $\eta$, the Dirac Lagrangian becomes
\begin{eqnarray}\label{Lstep1}
   {\cal L} &=& \bar\psi\,i\rlap{\,/}{D}\,\psi
    = (\bar\xi+\bar\eta)\,i\rlap{\,/}{D}\,(\xi+\eta) \nonumber\\
   &=& \bar\xi\,\frac{\rlap/\bar n}{2}\,in\cdot D\,\xi 
    + \bar\eta\,\frac{\rlap/n}{2}\,i\bar n\cdot D\,\eta 
    + \bar\xi\,i\rlap{\,/}{D}_\perp\,\eta 
    + \bar\eta\,i\rlap{\,/}{D}_\perp\,\xi \,. 
\end{eqnarray}
Here the covariant derivative $D$ still contains both the soft and the 
hard-collinear gluon fields. We now ``integrate out'' the small-component 
field $\eta$ by solving its equation of motion:
\begin{equation}\label{eta}
   \frac{\rlap/n}{2}\,i\bar n\cdot D\,\eta + i\rlap{\,/}{D}_\perp\,\xi = 0
   \quad \Rightarrow \quad
   \eta = - \frac{\rlap/\bar n}{2}\,\frac{1}{i\bar n\cdot D+i\epsilon}\,
    i\rlap{\,/}{D}_\perp\,\xi \,,
\end{equation}
where the ``$i\epsilon$'' prescription is arbitrary but must be fixed once 
and forever. Note that the result is highly non-local, involving an inverse
differential operator acting on the hard-collinear quark field.

It is instructive to compare this solution with the corresponding expression 
(\ref{Hv}) for the ``small'' components of the heavy-quark field in HQET.
In this case, the result was ``almost local'' in that the derivative in the 
denominator produces a power of the residual momentum $v\cdot k$, which is
much smaller than $2m_Q$. It was therefore possible to expand the expression
(\ref{Hv}) in an infinite series of local operators, and this produced the
terms in the (local) HQET Lagrangian. In the case of SCET such a local 
expansion is not possible. In order to proceed, it is useful to introduce
the Wilson line
\begin{equation}\label{firstW}
   W(x) = P\exp\left( i g_s \int_{-\infty}^0\!dt\,\bar n\cdot A(x+t\bar n)
   \right) ,
\end{equation}
where $A^\mu=A_{hc}^\mu+A_s^\mu$ still contains both soft and hard-collinear 
gluons. The ordering prescription is the same as in (\ref{Sdef}). In analogy
with (\ref{Srule}), this object obeys the relations
\begin{equation}
   W^\dagger\,i\bar n\cdot D\,W = i\bar n\cdot\partial \,, \qquad
   \frac{1}{i\bar n\cdot D+i\epsilon}
   = W\,\frac{1}{i\bar n\cdot\partial+i\epsilon}\,W^\dagger \,.
\end{equation}
This allows us to turn the inverse covariant derivative in (\ref{eta}) 
into an ordinary inverse derivative, i.e., an integral. The result is
\begin{eqnarray}
   \eta(x)
   &=& W(x)\,\frac{\rlap/\bar n}{2}\,
    \frac{(-1)}{i\bar n\cdot\partial+i\epsilon}\,\left( W^\dagger\,
    i\rlap{\,/}{D}_\perp\,\xi \right)(x) \nonumber\\
   &=& W(x)\,\frac{\rlap/\bar n}{2}\,i\int_{-\infty}^0\!dt
    \left( W^\dagger\,i\rlap{\,/}{D}_\perp\,\xi \right)(x+t\bar n) \,.
\end{eqnarray}
Inserting this into (\ref{Lstep1}) yields the non-local effective Lagrangian
\begin{equation}
   {\cal L} = \bar\xi(x)\,in\cdot D(x)\,\xi(x) 
   + \left( \bar\xi\,i\rlap{\,/}{D}_\perp W\right)(x)\,
   \frac{\rlap/\bar n}{2}\,i\int_{-\infty}^0\!dt
   \left( W^\dagger\,i\rlap{\,/}{D}_\perp\,\xi \right)(x+t\bar n) \,.
\end{equation}

At this stage our job is almost done. The remaining problem we need to 
deal with is that the above expression does not yet respect a proper power
counting, since terms of different orders in $\lambda$ are mixed up. In order 
to avoid double counting, it is imperative to expand all objects in the 
effective Lagrangian such that the power counting is consistent. For
instance, the different components of the covariant derivative acting on 
hard-collinear fields scale like
\begin{equation}
   iD^\mu = i\partial^\mu + g_s A_{hc}^\mu + g_s A_s^\mu
   \sim (\lambda,1,\lambda^{1/2}) + (\lambda,1,\lambda^{1/2})
   + (\lambda,\lambda,\lambda) \,.
\end{equation}
While for $n\cdot D$ all three contributions are of order $\lambda$, for
$\bar n\cdot D$ and $D_\perp$ the contributions involving the soft gluon 
field are power suppressed and should be neglected at leading order. 
Likewise, in the definition of the Wilson line $W$ in (\ref{firstW})
the contribution of the 
soft gluon field in the exponent is power suppressed and should be neglected. 
It follows that $W=W_{hc}+O(\lambda)$, where
\begin{equation}
   W_{hc}(x) = P\exp\left( i g_s \int_{-\infty}^0\!dt\,
   \bar n\cdot A_{hc}(x+t\bar n) \right) .
\end{equation}
Finally, in interactions with hard-collinear fields, soft fields must be
multi-pole expanded in order to ensure a proper power counting 
\cite{Beneke:2002ph}. To see why, consider as an example 
the phase factor associated with the coupling of a hard-collinear
field (incoming momentum $p_{hc}$) to a soft field (incoming momentum $p_s$), 
producing a hard-collinear field (outgoing momentum $p_{hc}'$):
\begin{equation}
   S_{\rm int} \ni \int d^4x\,\phi_{hc}'(x)\,\phi_{hc}(x)\,\phi_s(x)
   \sim \int d^4x\,e^{i(p_{hc}'-p_{hc}-p_s)\cdot x}\,\phi_{hc}'(0)\,
   \phi_{hc}(0)\,\phi_s(0) \,.
\end{equation}
The combined momentum in the exponent scales like a hard-collinear momentum,
since $(p_{hc}'-p_{hc}-p_s)^\mu\sim(\lambda,1,\lambda^{1/2})$. Consequently, 
$O(1)$ contributions to the action arise if 
$x^\mu\sim(1,\lambda^{-1},\lambda^{-1/2})$. However, in this case the phase
factor involving the soft momentum can be Taylor expanded:
\begin{equation}
   e^{-ip_s\cdot x} = e^{-ip_{s+}\cdot x_-} \left( 1
   - ip_{s\perp}\cdot x_\perp - ip_{s-}\cdot x_+ + \dots \right) ,
\end{equation}
where the terms inside the bracket scale like 1, $\sqrt{\lambda}$, and 
$\lambda$, respectively. It follows that in interactions with hard-collinear
fields we should expand soft fields as
\begin{equation}
   \phi_s(x) = \left( 1 + x_\perp\cdot\partial_\perp + \dots \right)
   \phi_s(x_-) \,.
\end{equation}
It is understood that the derivatives are evaluated before setting $x=x_-$ 
in the argument of the field $\phi_s$. At leading order, only the first term 
on the right-hand side contributes.

With all this insight, we are now prepared to write down the leading-power
terms in the SCET Lagrangian. The result is
\begin{eqnarray}
   {\cal L}_{\rm SCET}
   &=& \bar\xi(x)\,in\cdot D_{hc}(x)\,\xi(x) 
    + \bar\xi(x)\,g_s n\cdot A_s(x_-)\,\xi(x) \nonumber\\
   &&\mbox{}+ \Big( \bar\xi\,i\rlap{\,/}{D}_{hc\perp} W_{hc} \Big)(x)\,
    \frac{\rlap/\bar n}{2}\,i\int_{-\infty}^0\!dt
    \left( W_{hc}^\dagger\,i\rlap{\,/}{D}_{hc\perp}\,\xi \right)(x+t\bar n)
    \,.
\end{eqnarray}
It is easy to check that each term in this Lagrangian scales like 
$\lambda^2$, which when integrated with the measure $d^4x\sim\lambda^{-2}$ 
gives a leading, $O(1)$ contribution to the action. It is in principle 
straightforward to work out the higher-order corrections to the effective 
Lagrangian, but I will spare you the rather lengthy details and instead refer
to the literature \cite{Chay:2002vy,Beneke:2002ph}.

At leading order in power counting, there are no terms in the SCET Lagrangian 
containing interactions of soft quark fields with hard-collinear fields. 
There are, however, interactions among soft and hard-collinear gluons. Their 
explicit form at leading and subleading power can be found in 
\cite{Beneke:2002ph}. 
From the SCET Lagrangian, one can derive Feynman rules for soft and 
hard-collinear fields in the usual way \cite{Bauer:2001yt}. While these rules 
are more complicated that in HQET or even in full QCD, they nevertheless 
allow for a straightforward calculation of Feynman diagrams in SCET.

\subsubsection*{Residual gauge invariance}

The effective Lagrangian of SCET (including power corrections) is invariant, 
order by order in $\lambda$, under a set of residual hard-collinear and
soft gauge transformations, $U_{hc}$ and $U_s$, that preserve the 
scaling properties of the various fields 
\cite{Bauer:2001ct,Bauer:2001yt,Beneke:2002ph}. In analogy with 
(\ref{residualgauge}), one finds that under a soft gauge transformation 
\begin{eqnarray}
   h_v(x) &\to& U_s(x)\,h_v(x) \,, \nonumber\\
   q_s(x) &\to& U_s(x)\,q_s(x) \,, \nonumber\\
   A_s^\mu(x) &\to& U_s(x)\,A_s^\mu(x)\,U_s^\dagger(x)
    + \frac{i}{g_s}\,U_s(x)\,[\partial^\mu,U_s^\dagger(x)] \,, \nonumber\\
   \xi(x) &\to& U_s(x_-)\,\xi(x) \,, \nonumber\\
   A_{hc}^\mu(x) &\to& U_s(x_-)\,A_{hc}^\mu(x)\,U_s^\dagger(x_-) \,. 
\end{eqnarray}
Note the multi-pole expansion 
whenever soft fields are coupled to hard-collinear ones. Similarly, under a
hard-collinear gauge transformation
\begin{eqnarray}
   \xi(x) &\to& U_{hc}(x)\,\xi(x) \,, \nonumber\\
   n\cdot A_{hc}(x) &\to& U_{hc}(x)\,n\cdot A_{hc}(x)\,U_{hc}^\dagger(x)
    + \frac{i}{g_s}\,U_{hc}(x)\,[n\cdot D_s(x_-),U_{hc}^\dagger(x)] \,,
    \nonumber\\
   A_{hc}^\mu(x) &\to& U_{hc}(x)\,A_{hc}^\mu(x)\,U_{hc}^\dagger(x)
    + \frac{i}{g_s}\,U_{hc}(x)\,[\partial^\mu,U_{hc}^\dagger(x)] \,; \quad
    \mu\ne + \,, 
\end{eqnarray}
while all soft fields remain invariant. As before,
``soft'' and ``hard-collinear'' functions like $U_s(x)$ and $U_{hc}(x)$ can be 
defined via a restriction to soft and hard-collinear modes in their Fourier 
decompositions, respectively. 

\subsubsection*{Decoupling transformation}

At leading order in $\lambda$, soft gluons couple to hard-collinear fields
only via interactions involving $n\cdot A_s$. The reason is simply that this
is the only component of the soft gluon field which scales in the same way as
the corresponding component of the hard-collinear gluon field. In analogy
with our discussion for HQET, these interactions can be removed by a field
redefinition \cite{Bauer:2001yt}. Let us define new fields
\begin{eqnarray}\label{decoupl}
   \xi(x) &=& S_n(x_-)\,\xi^{(0)}(x) \,, \nonumber\\
   A_{hc}^\mu(x) &=& S_n(x_-)\,A_{hc}^{\mu(0)}(x)\,S_n^\dagger(x_-) ,
\end{eqnarray}
where
\begin{equation}
   S_n(x_-) = P\exp\left( i g_s \int_{-\infty}^0\!dt\,n\cdot A_s(x_-+tn)
   \right)
\end{equation}
is a light-like soft Wilson line along the $n$ direction. In terms of the new
fields, the SCET Lagrangian 
\begin{eqnarray}
   {\cal L}_{\rm SCET}
   &=& \bar\xi^{(0)}(x)\,in\cdot D_{hc}^{(0)}(x)\,\xi^{(0)}(x) \nonumber\\
   &&\mbox{}+ \left( \bar\xi^{(0)}\,i\rlap{\,/}{D}_{hc\perp} W_{hc}^{(0)}
    \right)(x)\,
    \frac{\rlap/\bar n}{2}\,i\int_{-\infty}^0\!dt
    \left( W_{hc}^{\dagger(0)}\,i\rlap{\,/}{D}_{hc\perp}^{(0)}\,\xi^{(0)}
    \right)(x+t\bar n)
\end{eqnarray}
has decoupled from soft gluons. As in the case of HQET, to see whether this
is useful we will have to study what happens to external operators containing
hard-collinear fields.

The decoupling transformation (\ref{decoupl}), and a similar transformation 
for the effective theory SCET-2 \cite{Becher:2003kh}, play an important 
role in proofs of soft-collinear factorization theorems, see e.g.\ 
\cite{Bauer:2001yt,Bauer:2001cu,Bosch:2003fc,Bosch:2004th,Becher:2005fg}.

\subsubsection*{Heavy-light currents}

\begin{figure}
\begin{center}
\includegraphics[width=0.65\textwidth]{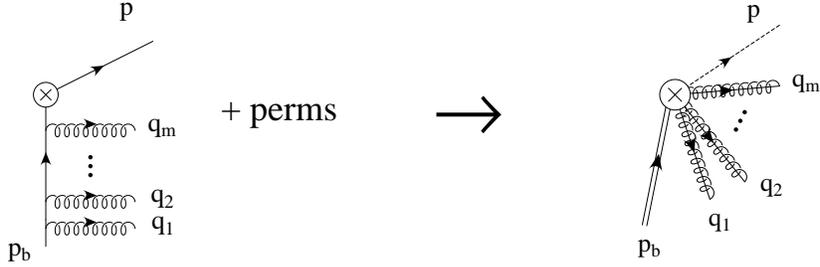} 
\end{center}
\vspace*{-0.5cm}
\caption{\label{fig:x}
Tree-level matching for heavy-light currents from QCD onto SCET. In the 
diagram on the right, the double line represents the heavy quark, the dashed
line the hard-collinear quark, and the crossed circle the effective 
current operator.
(Figure taken from \cite{Bauer:2000yr} with permission from the authors)}
\end{figure}

Weak-interaction processes are mediated by flavor-changing quark currents, 
which in QCD have the generic form $J=\bar\psi\,\Gamma\,Q$, where $Q$ is a
heavy quark and $\Gamma$ denotes some Dirac structure. An important questions
is what happens to such operators after matching onto SCET. The naive guess,
$J\to\bar\xi\,\Gamma\,h_v$ is not gauge invariant with respect to the soft 
and hard-collinear gauge transformations discussed above. Tree-level
matching shows that the correct answer (at tree level only!) is 
\cite{Bauer:2000yr}
\begin{equation}
   J(x)\to \left(\bar\xi\,W_{hc}\right)(x)\,\Gamma\,h_v(x_-) \,,
\end{equation}
which is gauge invariant. This matching is illustrated in Figure~\ref{fig:x}. 
Remarkably, the effective current operator in SCET contains an infinite 
number of vertices involving an arbitrary number of 
$\bar n\cdot A_{hc}\sim 1$ gluons, whose couplings are unsuppressed. All of 
these vertices are present at leading power in $\lambda$. 

Beyond tree level, the correct matching relation is yet more complicated. One
can show that the most general gauge-invariant form is \cite{Beneke:2002ph}
\begin{eqnarray}\label{Jmatch}
   J(x) &\to& \sum_i \int dt\,\widetilde C_i(t,\mu)
    \left(\bar\xi\,W_{hc}\right)(x+t\bar n)\,\Gamma_i\,h_v(x_-) \nonumber\\
   &=& \sum_i C_i(\bar n\cdot\bm{P}_{hc},\mu)
    \left(\bar\xi\,W_{hc}\right)(x)\,\Gamma_i\,h_v(x_-) \,.
\end{eqnarray}
Here $\Gamma_i$ are Dirac
matrices constructed with the help of the 4-vectors $v$, $n$, and $\bar n$, 
which must have the same quantum numbers as the original matrix $\Gamma$. The 
symbol $\bm{P}_{hc}$ represents the operator of total hard-collinear momentum, 
which by definition is the total momentum of all hard-collinear fields. In
the last step in (\ref{Jmatch}) we have used translational invariance. 
Finally, we have defined the Fourier-transformed coefficients
\begin{equation}
   C_i(\bar n\cdot p,\mu)
   = \int dt\,e^{i\bar n\cdot p t}\,\widetilde C_i(t,\mu) \,.
\end{equation}

Let us see what happens to the result (\ref{Jmatch}) when we apply the
decoupling transformations (\ref{hvdecoupl}) and 
(\ref{decoupl}). In terms of the new fields, we find
\begin{equation}
   J(x) = \sum_i C_i(\bar n\cdot\bm{P}_{hc},\mu)
   \big( \bar\xi^{(0)} W_{hc}^{(0)} \big)(x)\,\Gamma_i
   \left[ S_n^\dagger(x_-)\,S_v(x_-) \right] h_v^{(0)}(x_-) \,.
\end{equation}
Recall that the redefined heavy-quark field $h_v^{(0)}$ is sterile (it does not
couple to anything), while the redefined hard-collinear fields do not couple
to soft gluons. Nevertheless, similar to (\ref{newcurrent}), 
the soft gluon fields have
not disappeared from the final expression. Rather, they are once again
contained in a soft Wilson loop, $[S_n^\dagger(x_-)\,S_v(x_-)]$. The presence 
of a cusp at the point $x_-$ implies that the anomalous dimension of the 
effective current operators contains a term proportional to the cusp anomalous
dimension \cite{Becher:2003kh},
\begin{equation}\label{Jgamma}
   \gamma_{J_{\rm eff}} = \Gamma_{\rm cusp}(\alpha_s)\,
   \ln\frac{\bar n\cdot P_{hc}^{\rm tot}}{\mu} + \gamma'(\alpha_s) \,.
\end{equation}
The remaining
term $\gamma'$ results from the residual hard-collinear self interactions of
the composite field $\bar\xi^{(0)} W^{(0)}$ located at point $x$. 
The anomalous dimension $\gamma'$ (\ref{Jgamma}) is known at two-loop order 
\cite{Neubert:2004dd}, while $\Gamma_{\rm cusp}$ is known to three loops 
\cite{Moch:2004pa}. With the help of these results, it is
possible to resum Sudakov double and single logarithms at NLO in 
RG-improved perturbation theory (see \cite{Neubert:2004dd} for an explicit 
example).

\subsection*{Sample application: Factorization in \boldmath$B\to X_s\gamma$ 
decay\unboldmath}

As a final, important example I discuss how soft-collinear factorization 
works for the rare, inclusive decay $B\to X_s\gamma$, following the 
discussions in \cite{Bauer:2001yt,Bosch:2004th,Neubert:2004dd}. This will 
provide a prototype application of a two-step matching procedure. In the
first step, hard quantum fluctuations at scales $\mu\sim m_b$ or above are 
integrated out by matching the effective weak Lagrangian onto a bilocal 
operator in SCET. In the second step, fluctuations at the hard-collinear 
scale are integrated out by matching this bilocal SCET operator onto a 
bilocal operator in heavy-quark effective theory. While we will discuss 
this two-step matching at leading power in $\Lambda_{\rm QCD}/m_b$, the 
procedure can be extended systematically to higher orders in the heavy-quark 
expansion \cite{Lee:2004ja,Bosch:2004cb,Beneke:2004in}. 

The derivation of the factorization formula proceeds in several steps, which 
I can only sketch here. In 
the first step, we use the optical theorem to relate the total decay rate to
the imaginary part of the forward scattering amplitude:
\begin{equation}
   \Gamma(B\to X_s\gamma)\propto\mbox{Im} \int d^4x\,
   \langle B|\,T\,\{ {\cal L}_{\rm eff}^{b\to s\gamma}(x)\,
   {\cal L}_{\rm eff}^{b\to s\gamma}(0) \}\,|B\rangle \,.
\end{equation}
In the second step we integrate out hard fluctuations by matching the
effective weak Lagrangian ${\cal L}_{\rm eff}^{b\to s\gamma}$ given 
in (\ref{LeffW}) onto SCET 
operators containing a heavy-quark field $h_v$ and a hard-collinear field
$\xi$ describing the strange quark. At leading power all other fields are
hard and integrated out. The result is (setting $V_{ub}\to 0$ for simplicity)
\cite{Neubert:2004dd}
\begin{equation}
   {\cal L}_{\rm eff}^{b\to s\gamma}
   = \frac{G_F}{\sqrt2}\,V_{cb} V_{cs}^*\,\frac{e}{2\pi^2}\,E_\gamma\,
   \overline{m}_b(\mu)\,H_\gamma(\mu)
   \left(\bar\xi\,W_{hc}\right)(x)\,\rlap/\epsilon_\perp^*(q)\,(1-\gamma_5)\,
   h_v(x_-) + \dots \,,
\end{equation}
with
\begin{equation}
   H_\gamma(\mu) = C_{7\gamma}(\mu) - \frac13\,C_5(\mu) - C_6(\mu)
   + O(\alpha_s) \,.
\end{equation}
I won't bother writing down the $O(\alpha_s)$ corrections to the hard 
function, which have been calculated in the same paper.
In the third step, we integrate out hard-collinear fields by matching the 
two-point correlator in SCET onto bilocal operators in HQET. 
Before doing this, 
we decouple soft interactions from the hard-collinear fields by performing
the field redefinition (\ref{decoupl}). Then all hard-collinear fields are 
contained in the ``jet function'' \cite{Bauer:2001yt,Bosch:2004th}
\begin{equation}
   J(x)\equiv \langle 0|\,T\left\{
   \big( W_{hc}^{\dagger(0)}\,\xi^{(0)} \big)(x)
   \big( \bar\xi^{(0)}\,W_{hc}^{(0)} \big)(0) \right\} |0\rangle \,.
\end{equation}
Since the hard-collinear fields live at a perturbative scale of order 
$\mu_{hc}^2\sim E_\gamma\Lambda_{\rm QCD}$, the jet function can be 
calculated in perturbation theory. In fact, in the particular gauge 
$\bar n\cdot A_{hc}=0$ it is nothing but the quark propagator. 

After computing the jet function we are left with a $B$-meson forward matrix
element of soft heavy-quark fields in HQET, which is a non-perturbative 
object. The relevant matrix element is
\begin{equation}
   \langle B|\,\bar h_v(x_-)\,[x_-,0]\,h_v(0)\,|B\rangle \,,
\end{equation}
where the product $[x_-,0]\equiv S_n(x_-)\,S_n^\dagger(0)$ is a straight soft
Wilson line along the light-cone. This product 
arises from the field redefinition of the hard-collinear fields under the 
decoupling transformation (\ref{decoupl}). The Fourier transform of the 
above matrix element is called the shape function of the $B$ meson and
denoted by $S(\hat\omega,\mu)$ \cite{Neubert:1993ch}. It is
a conventional parton distribution function, defined however in the context
of HQET. The shape function is a simple, non-perturbative function describing
the internal dynamics of the $B$ meson.

We are now ready to collect the result for the inclusive decay rate or, more
precisely, for the photon-energy spectrum in the region of large photon 
energy, where the collinear expansion is justified. After Fourier 
transforming to momentum space, one obtains
\cite{Neubert:2004dd,Korchemsky:1994jb}
\begin{eqnarray}\label{QCDF}
   \frac{d\Gamma(B\to X_s\gamma)}{dE_\gamma}
   &=& \frac{G_F^2\alpha}{2\pi^4}\,|V_{cb} V_{cs}^*|^2\,
    \overline{m}_b^2(\mu)\,|H_\gamma(\mu)|^2\,E_\gamma^3 \nonumber\\
   &\times& \int\limits_0^{M_B-2E_\gamma}\!\!d\hat\omega\,
    J(m_b(M_B-2E_\gamma-\hat\omega),\mu)\,S(\hat\omega,\mu)
    + \dots \,,
\end{eqnarray}
where the ellipses represent power-suppressed terms. This formula achieves a 
complete factorization (separation) of short- and long-distance physics. The
short-distance physics resides in the perturbative hard function $H_\gamma$ 
and in the perturbative jet function $J$. The characteristic scales 
associated with these objects are $\mu_h\sim E_\gamma\sim m_b$ and 
$\mu_{hc}\sim\sqrt{E_\gamma\Lambda_{\rm QCD}}$, respectively. The 
hadronic physics is parameterized by the non-perturbative shape function.
Large logarithms associated with the various scales can be resummed to all
orders in perturbation theory by solving the RG evolution equations for the
hard, jet, and shape functions. This is discussed in detail in the 
literature \cite{Neubert:2004dd,Neubert:2005nt}.

Relation (\ref{QCDF}) is a beautiful example of a QCD factorization formula.
The technology of SCET and HQET, and of EFT in general, have made the 
derivation of such factorization statements a bit more straightforward then it 
was previously. Similar relations can be established for many other processes
in $B$ physics \cite{Bauer:2001cu,Bosch:2003fc,Becher:2005fg,Beneke:1999br}
and elsewhere (e.g., in DIS and jet physics \cite{Bauer:2002nz}). In many 
cases factorization is a property of the amplitude in the large-energy limit, 
but non-factorizable corrections arise at subleading order in the power 
expansion. Prominent examples are exclusive $B$ decays such as $B\to\pi\pi$ 
or $B\to K^*\gamma$, which play an important role in the physics program 
of the $B$-factories. In the case of inclusive $B$ decays considered in this
lecture, a factorization formula can be established at every order in 
the $1/E$ expansion \cite{Lee:2004ja,Bosch:2004cb,Beneke:2004in}. 
Inclusive $B$-decay distributions are therefore examples of a small
class of observables for which a systematic, field-theoretic description of
power corrections can be given in terms of non-local string operators 
(operators whose component fields are separated in space-time), in 
generalization of the local OPE valid in Euclidean space.

\subsubsection*{Acknowledgments}

I am grateful to the organizers of TASI 2004 for the invitation to present 
these lectures and for a flawless organization of the school. I am grateful 
to the students for their interest in the material presented here and for 
many lively discussions. I am particularly grateful to John Terning for his
patience in waiting for these notes to appear. I like to thank the 
Institute for Theoretical Physics at the University of Heidelberg, Germany, 
for its hospitality during the completion of these notes. Work on this project 
was supported in part by a Research Award of the Alexander von Humboldt 
Foundation, and by the National Science Foundation under Grant PHY-0355005.

\end{document}